\documentclass[aps, twocolumn, showpacs, preprintnumbers, amsmath, amssymb, prb]{revtex4-2}
\usepackage{amsfonts}
\usepackage{mathrsfs}
\usepackage{amsmath}% needed for subequations
\usepackage{color}
\usepackage{natbib}
\usepackage{textcomp}
\usepackage{graphicx}
\usepackage{enumerate}
\usepackage{bm}% bold maths
\usepackage{amssymb}
\usepackage{xspace}
\usepackage{epstopdf}
\usepackage{dcolumn}% Align table columns on decimal point
\usepackage{longtable}
\usepackage{subfigure}
\usepackage{multirow}
\usepackage{makecell}
\usepackage{exscale}
\usepackage{relsize}
\usepackage[colorlinks=true, letterpaper=true, pdfstartview=FitV, linkcolor=blue, citecolor=blue, urlcolor=blue]{hyperref}
\usepackage{float}
\usepackage[figuresright]{rotating}
\usepackage{longtable}
\usepackage{booktabs}
\usepackage{tabularx}
\usepackage{hyperref}
\usepackage{braket}

\makeatletter

\newcommand{\Rmnum}[1]{\expandafter\@slowromancap\romannumeral #1@}
\makeatother

\begin{document}
	\title{Effect of disorder on Berry curvature and quantum metric
		in two-band gapped graphene}
	
	\author{Ze Liu$^{1}$}
	\altaffiliation{The authors contribute equally.}
	\author{Zhi-Fan Zhang$^{1}$}
	\altaffiliation{The authors contribute equally.}
	\author{Zhen-Gang Zhu$^{1,2,3}$}
	\email{zgzhu@ucas.ac.cn}
	
	\author{Gang Su$^{3}$}
	\email{gsu@ucas.ac.cn}
	
	\affiliation{
		$^{1}$ School of Physical Sciences, University of Chinese Academy of Sciences, Beijing 100049, China. \\
		$^{2}$ School of Electronic, Electrical and Communication Engineering, University of Chinese Academy of Sciences, Beijing 100049, China.\\
		$^{3}$ Kavli Institute for Theoretical Sciences, University of Chinese Academy of Sciences, Beijing 100190, China.
	}
	\date{\today}
	
	\begin{abstract}
		The geometric properties of parameter space are mostly described by Berry curvature and quantum metric, which are the imaginary and real part of quantum geometric tensor, respectively. In this work, we calculate the dressed Berry curvature and quantum metric containing eight Feynman diagrams, which are proportional to the leading-order of the concentration of impurities.
		For a two-band gapped graphene model, we find the disorder does not break the original symmetry but decrease (increase) the absolute value of Berry curvature and quantum metric in conduction (valence) band. 
		We show how impurities affect the Berry curvature and quantum metric, deepening our understanding of the impurity effect on the electron transport properties in two-band gapped graphene.
	\end{abstract}
	\maketitle
	\section{INTRODUCTION}
	As the frontier of condensed matter physics, the geometric properties of quantum states are of great significance in many areas.
	One of the most general description of the evolution of a cell-periodic Bloch state under the variation of a vector parameter $\mathbf{k}$ is the quantum geometric tensor (QGT) proposed by Provost and Vallee \cite{Provost1980}.
	Recently, many theories and experiments have shown that QGT plays a peculiar and important role in electron transports.
	Its imaginary (anti-symmetric) part, i.e. Berry curvature (BC), a quantity manifested in many transport properties and observed in many materials, can be regarded as an emergent gauge field in the parameter space \cite{1984a,Berry1985,Xiao2010}.
	As we know, the integration of BC over the momentum space leads to the famous topological invariants (TKNN number or Chern number) in quantum anomalous Hall effect \cite{Shen2017,Thouless1982,Hasan2010,Qi2011,Fu2006,Fu2007}.
	Beyond the linear response theory, BC has an important role in nonlinear response effects \cite{Sodemann2015,Zhang2023,Gao2019,Ortix2021,Saha2023,Zhao2023,Nishijima2023,Zhang2021,Wang2022,Zhang2023a}.
	The quantum metric (QM), the real (symmetric) part of QGT,  is another important geometric quantity \cite{Chruscinski2004,Resta2011,AustrichOlivares2022,Gonzalez2019,Juarez2023,Ding2022}.
	It is worth noting that QM also has a non-negligible contribution to the intrinsic second-order Hall effects \cite{Liu2021,Wang2021,Gao2023,Wang2023,Das2023}, which is due to the fact that QM is expected to appear in the anomalous velocity of electrons proportional to the square of external electric field \cite{Gao2014} and has an equivalent value to the anomalous velocity induced by BC.
	What's more, many other physical phenomena related to QM have been proposed \cite{Chen2024,Smith2022,Piechon2016,Palumbo2018,Ozawa2021,Wang2023a,Wang2024}, such as the role of QM in governing superconductivity and superfluidity in platforms including graphene \cite{Rossi2021,Toermae2022}.
	As an illustration, Chen and Law proposed a Ginzburg-Landau theory from a microscopic flat-band Hamiltonian to show that the superconducting coherence length is determined directly by the QM \cite{Chen2024}.
	%
	%Further, some experiments have been done to confirm the nonlinear quantum transports induced by the QM \cite{Gao2023,Wang2023,}.
	%
	Generally speaking, the BC and QM  are both fundamental quantities for understanding many new topological effects. %, either  known or unknown.

	The BC and QM reflect the topology of many-particle systems and regulate particles' dynamics. Thus a quite fundamental question may raise whether the dynamical properties and many-body interaction can mutually modify the topological properties of the system.
	%
	%As we know, there are many many-body interactions to be considered, such as disorder, electron-electron, or electron-phonon interactions.
	%
	To be more specific, a precise discussion how the interaction affects BC and QM is urgently needed.
	Intriguingly, Wei Chen has provided a formalism for the BC and QM \cite{Chen2022}, which are applicable to realistic gapped materials at finite temperature and in the presence of many-body interactions.
	They introduce the spectral functions to characterize the BC and QM, and discuss how many-body interactions influence the shape of these spectral functions.
	Nevertheless, they obtain the leading order of impurity scattering which dose not contribute to the QM or BC in the Chern insulator because the integral of the spectral function is equal to zero.
	From another point of view, numerous theoretical and experimental results have proved that impurities can affect electron transport \cite{Mucciolo2010,Vojta1998,Wang2013,Hu2008,Du2019,Du2021,Atencia2023,Qiang2023}.
	As quantum transport is very closely related to the BC and QM, the role of disorder on them should be explored.

	In this work, we study this topic using eight Feynman diagrams containing the first-order contributions of the impurity with the help of the first Born approximation (FBA) and vertex correction.
	We specifically study  a two-band gapped graphene model, and find that the disorder contribute a certain correction, named as dressed BC and QM. %whose values is about $100$ times smaller than BC and QM.
	%
	%More importantly, the symmetry in $\mathbf{k}$ space is all broken by the dressed BC and QM, which may cause some unknown effects.
	Our calculation and results provide key information for understanding the influence of impurity interaction on BC and QM.

	The structure of this paper is organized as follows. In Sec. \ref{sec2}, we recall the definition and relations of QGT, BC and QM at zero temperature and non-interaction limit and the dressed BC and QM with many body interactions at finite temperatures.
	In Sec. \ref{sec3}, we exploit the eight Feynman diagrams about side jump scattering to explain how to calculate  the effect of disorder using the FBA and vertex correction. Next we discuss the results in a two-band gapped graphene model in Sec. \ref{sec4}. Finally we summarize and discuss our results in Sec. \ref{sec5}.

	\section{QUANTUM GEOMETRIC TENSOR}
	\label{sec2}
	%\subsection{BC and QM}
	The QGT of $n$-th band in the $\mathbf{k}$ space can be written as $\text{T}_\mathbf{k}^n$, whose imaginary (anti-symmetric) part is Berry curvature (BC), $\Omega _{\mu \nu ,\mathbf{k}}^n=-2\mathrm{Im}\text{T}_{\mu \nu ,\mathbf{k}}^n$, and real (symmetric) part is quantum metric (QM) tensor, $g_{ \mu \nu ,\mathbf{k}}^n=\mathrm{Re} \text{T}_{\mu \nu ,\mathbf{k}} ^n$ \cite{Provost1980,Piechon2016},
	%describe the evolution of a cell-periodic Bloch state under the variation of a vector parameter $k$,
	\begin{equation}
		\text{T}_{\mu \nu ,\mathbf{k}}^n=\left. \langle \partial _{\mu} u_\mathbf{k}^n \left| 1-\mathscr{P}^n(\mathbf{k}) \right|\partial _{\nu}u_\mathbf{k}^n \right. \rangle =g_{\mu \nu ,\mathbf{k}}^n+i\frac{\Omega _{\mu \nu ,\mathbf{k}}^n}{2},
		\label{eq1}
	\end{equation}
	where $\mu, \nu \in (x,y,z)$, $\partial _{\mu}=\partial_{k^{\mu}}$, $u^n_k$ is the periodic part of the Bloch wave function of energy band $n$ with crystal momentum $\mathbf{k}$, $|u_{\mathbf{k} }^n \rangle$ is a ket vector using the Dirac notation \cite{Greiner2000}, and $\mathscr{P} ^n\left( \mathbf{k} \right) =|u_\mathbf{k}^n\rangle \langle u_\mathbf{k}^n|$ is the projector on the band $n $. Quantum metric characterizes a distance in Hilbert space, defined as
	\begin{equation}
		ds^{2,n}=1-\left| \left. \langle u_{\mathbf{k}}^n| u_{\mathbf{k}+d\mathbf{k}}^n\right. \rangle \right|^2=g_{\mu \nu ,\mathbf{k}}^ndk^{\mu}dk^{\nu},
		\label{eq2}
	\end{equation}
	and the Berry curvature tensor is a gauge-field tensor derived from the Berry vector potential, $ \mathcal{A}_{\mathbf{k}}^{n}=i \langle u_\mathbf{k}^n|\nabla_\mathbf{k} |u_\mathbf{k}^n \rangle $, in analogy to electrodynamics \cite{Xiao2010,Shen2017},
	\begin{equation}
		\Omega _{\mathbf{k}}^n=\nabla \times  \mathcal{A}_\mathbf{k}^{n}.
		\label{eq3}
	\end{equation}
	%\subsection{BC and QM}
	
	For the sake of discussing many-body interactions perturbatively, the dressed Berry curvature and quantum metric at finite temperature can be proposed using the linear response theory. We now consider the application of an external electric field $E^\mu $, and the total Hamiltonian can be written as
	\begin{equation}
		\mathcal{H}^{\text{tot}}=\mathcal{H}_\mathbf{k}+\delta \mathcal{H}_\mathbf{k}=\mathcal{H}_\mathbf{k} -i q E^{\mu} \partial_{\mu},
		\label{eq4}
	\end{equation}
	where $ q$  indicates the charge of the particle,  $\mathcal{H}_\mathbf{k} $ is the unperturbed periodic Bloch Hamiltonian with $\mathcal{H}_\mathbf{k}  |u_\mathbf{k}^n \rangle =\varepsilon_\mathbf{k}^n  |u_\mathbf{k}^n \rangle$, and the perturbed Hamiltonian $\delta \mathcal{H}_\mathbf{k}$ is described by the dipole energy. Considering the second-quantization by using the creation (annihilation) operator $c^\dagger_{n,\mathbf{k}} (c_{n,\mathbf{k}} )$, the Hamiltonian of the whole system is  $\mathcal{H}_\mathbf{k}^{\text{tot}}=\mathcal{H}_\mathbf{k}+\delta \mathcal{H}_\mathbf{k} = \mathcal{H}_\mathbf{k}-qE^\mu \sum_n U_{\mu ,\mathbf{k}}^n $, where the charge polarization operator is $U_{\mu ,\mathbf{k}}^n = \sum_{n'}  \mathcal{A} ^{nn'}_{\mu ,\mathbf{k}} c^\dagger_{n,\mathbf{k}}  c_{n',\mathbf{k}} $. It evolves with time according to $U^n _{\mu ,\mathbf{k}}(t) = e^{iH_\mathbf{k}^{\text{tot}}t}U^n _{\mu ,\mathbf{k}}  e^{-i H_\mathbf{k}^{\text{tot}}t}$, with  Berry connection $\mathcal{A} ^{nn'}_{\mu ,\mathbf{k}}$ no dynamics. Thus, the thermodynamic average \cite{Bruss2003, Mahan2013} of the charge polarization operator is
	\begin{equation}
		\left. \langle U^n _{\mu ,\mathbf{k}}(t)\right. \rangle =\chi^n  _{\mu \nu ,\mathbf{k}}(t)qE^\nu (t),
	\end{equation}
	where $\chi^n  _{\mu \nu ,\mathbf{k}}(t)$ is the susceptibility and the driving electric field is $E^\nu (t)=E^\nu e^ {- i \omega t}$.
	Furthermore, the Matsubara version of the susceptibility is calculated by
	\begin{equation}
		\chi^n  _{\mu v,\mathbf{k}}(i\omega ) =-\int_0^{\beta}{d}\tau e^{i\omega \tau}\left. \langle T_{\tau}U^n_{\mu ,\mathbf{k}}(\tau)U^{n'}_{\mu ,\mathbf{k}}(0) \right. \rangle .
		\label{eq6}
	\end{equation}
	According to the linear response theory \cite{Mahan2013}, we discuss that the  $\chi^n  _{\mu v,\mathbf{k}}$ is proportional to the susceptibility of the electric dipole energy driven by the external electric field, which reflects the magnitude of the electric dipole energy caused by the electric field, and has very important physical significance.
	
	Based on the susceptibility, the dressed QGT, BC and QM are obtained as \cite{Chen2022}
	\begin{equation}
		\begin{aligned}
			T_{\mu v,\mathbf{k}}^{\text{d},n} &=\frac{1}{2\pi}\int_0^{\omega}{\left[ i\chi _{\mu v,\mathbf{k}}^n(\omega )-i\chi _{\nu \mu ,\mathbf{k}}^{n*}(\omega ) \right] d\omega ,}
			\\
			g_{\mu \nu ,\mathbf{k}}^{\text{d},n} &=-\frac{1}{2\pi}\int_0^{\omega}{\mathrm{Im}\left[ \chi ^n_{\mu v,\mathbf{k}}(\omega )+\chi^n _{\nu \mu ,\mathbf{k}}(\omega ) \right] d\omega ,}
			\\
			\Omega _{\mu v,\mathbf{k}}^{\text{d},n} &=-\frac{1}{\pi}\int_0^{\omega}{\mathrm{Re}\left[ \chi ^n_{\mu v,\mathbf{k}}(\omega )-\chi^n _{\nu \mu ,\mathbf{k}}(\omega ) \right] d\omega},
		\end{aligned}
		\label{eq7}
	\end{equation}
	where the superscript $\text{d}$ indicates these quantities are dressed by impurity scattering.
	%In passing, we can repeat the expression of BC and QM in Eq. (\ref{eq1}-\ref{eq3}) in the zero temperature and non-interacting limit in the Appendix \ref{AppendixA}.

	\section{THE EFFECTS OF DISORDER}
	\label{sec3}
	To study the effects of disorder in detail, we consider an elastic scattering \cite{Du2019}, i.e., $V(\mathbf{r})=\sigma_0 \sum_i u_i \delta(\mathbf{r}-\mathbf{R}_i)$, where $\sigma_0$ has the same dimension with the Hamiltonian, the disorder is modeled as $\delta$-function scatters with a random distribution $\{\mathbf{R}_i\}$, $u_i$ indicates the strength of disorder and $\sigma_0$ is a unit matrix.
	
\begin{figure}[tb]
		\includegraphics[width=1\linewidth]{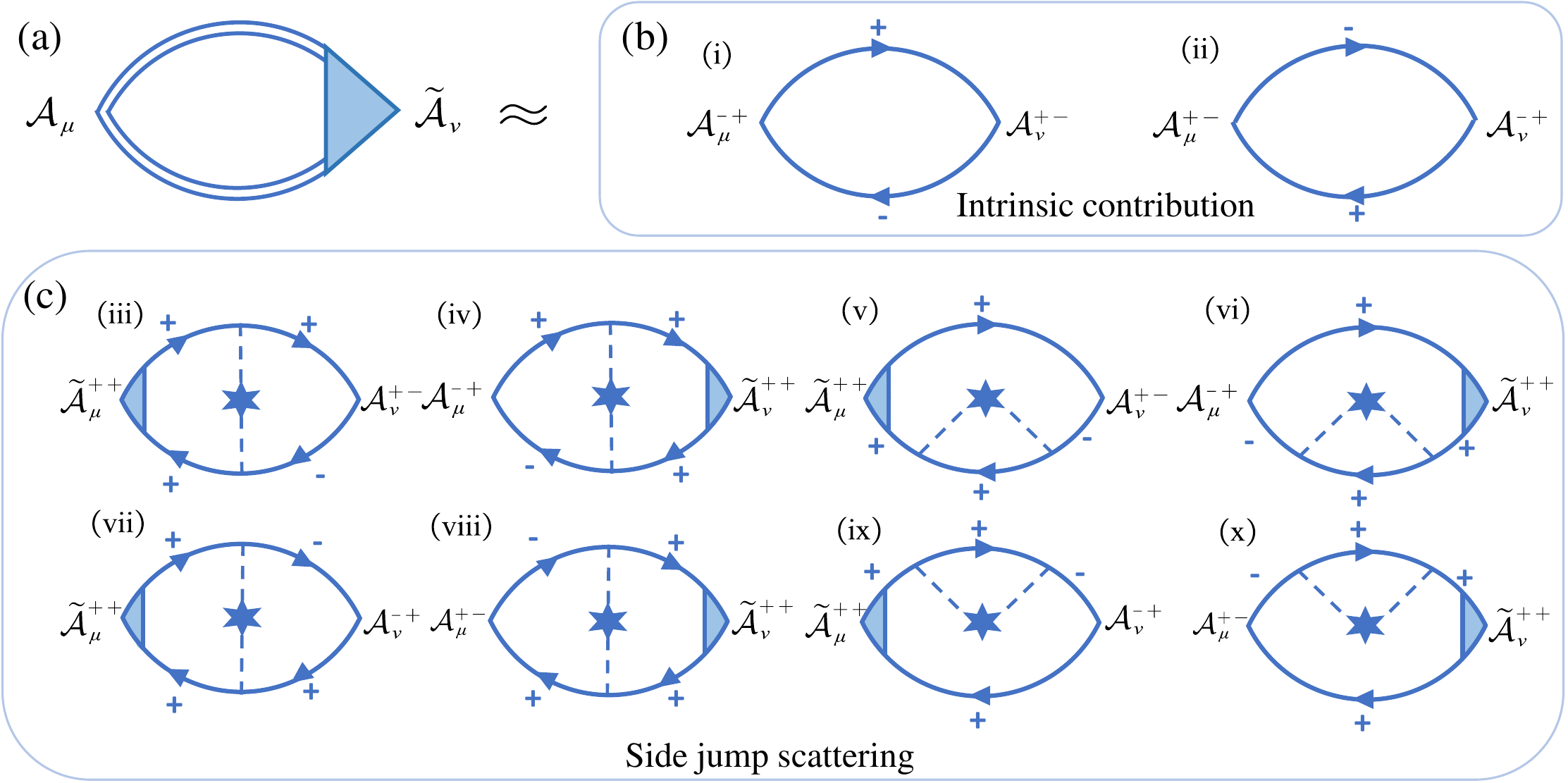}
		\caption{Feynman diagrams for susceptibility (a). Under FBA and vertex correction, these diagrams are classed into (b) intrinsic contribution and (c) extrinsic side jump scattering. (i) and (iii-vi) are for valence band while (ii) and (vii-x) are for conduction band. The double and single solid lines stand for the full and free-particles Matsubara Green's function, where arrows to the right and left indicate  retarded and advanced Green's function, respectively. The Eigen bands of a generic two-band model are labeled as $\pm$. The dash-star-dashed lines represent the disorder scattering. The light blue shadow represents the corrected vertices.}
		\label{fig1}
	\end{figure}

	Feynman diagrams of the susceptibility $ \chi _{\mu \nu,k}^n(\omega )$ is drew according to Eq. (\ref{eq6}), which is a two-particle Green's function with two operators at each of the external vertices \cite{Bruss2003} as shown in Fig. \ref{fig1}(a),
	\begin{equation}
		\chi^n_{\mu \nu ,\mathbf{k}}(\omega )=\sum _{m}\mathcal{A} _{\mu,\mathbf{k}}^{nm}\tilde{\mathcal{A}} _{\nu,\mathbf{k}}^{mn}\frac{1}{\beta}\sum_{ip}\mathscr{G} _{\mathbf{k}}^{m}(ip+i\omega )\mathscr{G}_{\mathbf{k}}^{n}(ip),
		\label{eq8}
	\end{equation}
	where $1/\beta = k_BT$, $k_B$ represents the Boltzmann constant, $T$ is the temperature,  $n,m$ show different energy bands, $\tilde{\mathcal{A}} _{\nu,\mathbf{k}}^{mn}$ represents the dressed vertices corrected by impurities, which we will discuss below. And $\mathscr{G} _{\mathbf{k}}^{}(ip)=\frac{1}{ip -\varepsilon _{\mathbf{k}}^{}-\Sigma_\mathbf{k}}$ is the full Green's function, where $ \Sigma_\mathbf{k}$ is the self-energy function for electrons.
	
	According to Eq. (\ref{eq8}), we have retained several diagrams of the lager impurity contributions drawn in Fig. \ref{fig1}. Diagrams (b) are intrinsic contributions, not related to the impurity scattering.
	By expanding the double-line full Green's function with the aid of FBA and dressed vertices, we can obtain eight diagrams [Fig. \ref{fig1}(c)] showing the effect of first-order disorder scattering, in which (iii)-(vi) are for $+$ band and (vii)-(x) are for $-$ band.
	According to these diagrams, the susceptibility in Eq. (\ref{eq8}) can be calculated, and the influence of the impurities on BC and QM can be further studied.
	%The detailed calculations of diagrams about side jump scattering in Fig. \ref{fig1} have been written in the Appendix \ref{AppendixB}.
	
	%%%%%%%%%%%%%%%%%%%%%%%%%%%%%%%%%%%%%%%%%%%%%%%%%%%%%%%%%%%%%%%%%%%%%%%%%%%%%%%%%%%%%%%%%%%%%%%%%%%%%%%%%%%%%%%%%%%%%%
	%%%%%%%%%%%%%%%%%%%%%%%%%%%%%%%%%%%%%%%%%%%%%%%%%%%%%%%%%%%%%%%%%%%%%%%%%%%%%%%%%%%%%%%%%%%%%%%%%%%%%%%%%%%%%%%%%%%%%%
	
	\subsection{The summation on Matsubara frequencies in Feynman diagrams}
	
	In order to derive the susceptibility, firstly we need to calculate the summation on Matsubara frequencies of Green's function in the Feynman diagrams.
	Following chapter 3 of Mahan \cite{Mahan2013}, we work with Matsubara frequencies, while allow the evaluation of the integrals with straightforward contour integral techniques. These details are shown in the Appendix \ref{AppendixA}.
	
	\subsection{The self-energy and relaxation time}
	Further, we need to solve for the self-energy and relaxation time of electrons. Herein, we define the free-particle Green's function is $G (ip)=\frac{1}{ip+i\omega -\varepsilon _{\mathbf{k}}}$ and the Dyson equation for  full Green's function $ \mathscr{G}$ can be written as $\mathscr{G}= G+G\Sigma \mathscr{G}$. Upon averaging disorders, the self-energy equation in the form of Dyson's equation is \cite{Sinitsyn2007}
	\begin{equation}
		\Sigma_\mathbf{k}=\langle V\rangle_{\text{imp}} +\langle VGV \rangle_{\text{imp}}+\langle VGVGV \rangle_{\text{imp}} + \cdots.
	\end{equation}
	To get the leading order of disorder, we have
	\begin{equation}
		\Sigma ^{\pm}=\langle V_{\mathbf{k}\mathbf{k}^{\prime}}^{\pm +}V_{\mathbf{k}^{\prime} \mathbf{k}}^{+\pm}\rangle_{\text{imp}}G^{+}+\langle V_{\mathbf{k}\mathbf{k}^{\prime}}^{\pm -}V_{\mathbf{k}^{\prime} \mathbf{k}}^{-\pm}\rangle_{\text{imp}}G^{-},
	\end{equation}
	where the eigenstates of a general two-band model are labeled as $n=\pm$ and $V^{\pm\pm}_{\mathbf{k}\mathbf{k}'}=\langle\pm,\mathbf{k}|V(\mathbf{r})|\pm,\mathbf{k'}\rangle$ is the scattering matrix element of single impurity potential after Fourier transformation. $\langle
	\cdots \rangle_{\text{imp}}$ means the average of disorder, satisfying $\langle u_i \rangle_{\text{imp}} =0, \langle u_i^2 \rangle_{\text{imp}} = u_0^2 \neq 0$, and the average concentration of impurities is $n_i$\cite{Koshino2006}. Mathematically, impurity averaging is achieved by summing over all phase-independent coherent subsystems and dividing by their number. But due to the random distribution of impurities, this average is the same as an average over the locations of all impurities within a single subsystem\cite{Bruss2003}.  Furthermore,  the relaxation time can be expressed as \cite{Mahan2013}
	\begin{equation}
		\frac{1}{\tau_\mathbf{k}}=-\frac{2}{\hbar}\text{Im}\Sigma_\mathbf{k}.
	\end{equation}

	\subsection{The dressed vertex}
	%%%%%%%%%%%%%%%%%%%%%%%%%%%%%%%%%%%%%%%%%%%%%%%%%%%%%%%%%%%%%%%%%%%%%%%%%%%%%%%%%%%%%%%%%%%%%%%%%%%%%%%%%%%%%%%%%%%%%%
	\begin{figure}[tbh]
		\includegraphics[width=0.6\linewidth]{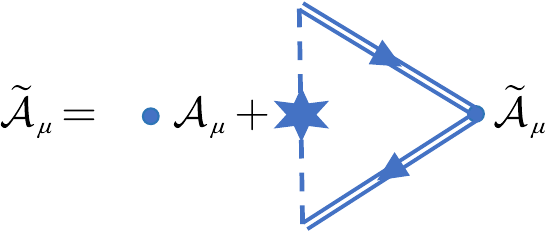}
		\caption{The Dyson equation for the dressed vertices. The double solid lines stand for the full  Matsubara Green's function. The dash-star-dashed line represents the disorder scattering. }
		\label{fig2}
	\end{figure}
	%%%%%%%%%%%%%%%%%%%%%%%%%%%%%%%%%%%%%%%%%%%%%%%%%%%%%%%%%%%%%%%%%%%%%%%%%%%%%%%%%%%%%%%%%%%%%%%%%%%%%%%%%%%%%%%%%%%%%%

	Shown in Fig. \ref{fig2}, the Dyson equation for dressed vertices  is written as \cite{Sinitsyn2007,Bruss2003}
	\begin{equation}
		\tilde{\mathcal{A}} _{\mu ,\mathbf{k}}^{nn} =\mathcal{A}^{nn}_{\mu, \mathbf{k}}+\sum_{\mathbf{k}''}\langle V^{nn}_{\mathbf{k}\mathbf{k}''}V^{nn}_{\mathbf{k}''\mathbf{k} }\rangle_{\text{imp}}\mathscr{G}^{R,n}_{\mathbf{k}''} \mathscr{G}^{A,n}_{\mathbf{k}''}
		\tilde{\mathcal{A}} _{\mu ,\mathbf{k}''}^{nn},
		\label{eq12}
	\end{equation}
	where the first term  free of impurities is the Berry connection without interaction.
	As we know, in vertex correction, every dash-star-dashed line that connects retarded and advanced Green's function provides a factor $n_i u_0^2$; while the Green's function product $\mathscr{G}^R\mathscr{G}^A$ is proportional $ (n_i u_0^2)^{-1}$ \cite{Mahan2013}.
	The overall vertex correction is independent of the concentration of impurities $n_i$. Thus, all the diagrams [Fig. \ref{fig1}(iii)-(x)] are the leading order of the concentration of impurities $n_i$ representing only the side jump scattering. Although higher order effects, i.e. skewing scattering, can also contribute to these geometric quantities, we have neglected them in this work.

	%%%%%%%%%%%%%%%%%%%%%%%%%%%%%%%%%%%%%%%%%%%%%%%%%%%%%%%%%%%%%%%%%%%%%%%%%%%%%%%%%%%%%%%%%%%%%%%%%%%%%%%%%%%%%%%%%%%%%%%%%%%
	\begin{figure}[tb]
		\includegraphics[width=1.1\linewidth]{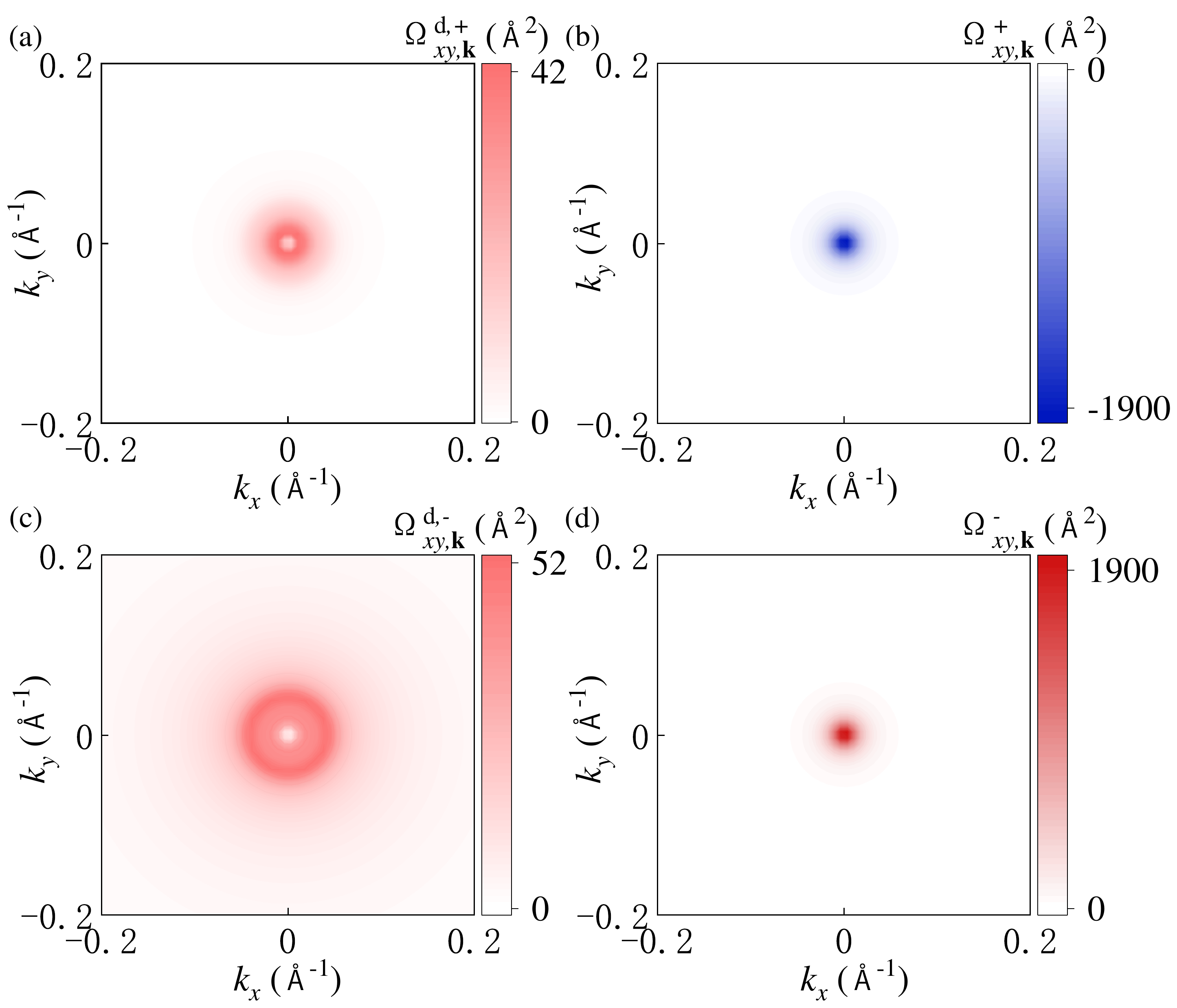}
		\caption{The 2D contour for off-diagonal component of (a),(c) dressed BC $\Omega_{xy,\mathbf{k}}^{\text{d}}$ and (b),(d) $ \Omega_{xy,\mathbf{k}}$ BC. (a)-(b) are for the $+$ band while (c)-(d) are for the $-$ band. The parameters are $ \Delta =0.2 $ eV \cite{Zhou2007}, $n_i u_0^2 L^2 = 10$ $\text{eV}^2 \text{\AA}^2$, $v_D = 10^6$ m/s, $\varepsilon_f = 0.3$ eV.}
		\label{fig3}
	\end{figure}
	%%%%%%%%%%%%%%%%%%%%%%%%%%%%%%%%%%%%%%%%%%%%%%%%%%%%%%%%%%%%%%%%%%%%%%%%%%%%%%%%%%%%%%%%%%%%%%%%%%%%%%%%%%%%%%%%%%%%%%%%%%%

	\section{TWO-BAND GAPPED GRAPHENE MODEL} \label{sec4}
	
	To make our discussion more specifically, we consider a two-band gapped graphene model \cite{Srivastava2015},
	\begin{equation}
		H=\\
		\begin{aligned}
			\begin{bmatrix}
				\frac{\Delta}{2} & v_D\hbar (\tau k_x -i k_y ) \\
				v_D\hbar (\tau k_x +i k_y ) & -\frac{\Delta}{2}
			\end{bmatrix},
		\end{aligned}
	\end{equation}
	where $\Delta$ is the gap, $\tau = \pm1$ are valley indexes, $v_D$ is the Dirac velocity, and $\varepsilon_\mathbf{k}^{\pm}=\pm \sqrt{v_D^2\hbar^2\mathbf{k}^2+\left(\frac{\Delta}{2}\right)^2}$. The BC is derived \cite{Xiao2010,Smith2022}
	\begin{equation}
		\Omega_{xy,\mathbf{k}}^\pm =\mp \frac{v_{D}^{2} \hbar^2 \Delta}{4\left[v_{D}^{2}\hbar^2\mathbf{k}^{2}+(\Delta / 2)^{2}\right]^{3 / 2}},
	\end{equation}
	and the four components of QM are derived as
	\begin{equation}
		g_{\mu \nu ,\mathbf{k}}^\pm=\frac{v_{D}^{2}\hbar^2\left\{ \left[ v_{D}^{2}\hbar^2\mathbf{k}^2+(\Delta /2)^2 \right] \delta _{\mu \nu  }-v_{D}^{2}\hbar^2k_\mu k_\nu  \right\}}{4\left[ v_{D}^{2}\hbar^2\mathbf{k}^2+(\Delta /2)^2 \right] ^2}.
	\end{equation}
	Motivated by the above discussions in Eqs. (\ref{eq8})-(\ref{eq12}) and considered only the disorder average proportional to the leading order of the
	concentration of impurities $n_i$, we can get self-energy and relaxation time (see Appendix \ref{AppendixB1}, \ref{AppendixB2})
	\begin{equation}
		\begin{aligned}
			&\Sigma^{ \pm}=-\frac{n_i u_0^2L^2}{4\pi v_D^2\hbar^2}\left(\varepsilon_\mathbf{k} \pm \frac{\Delta^2}{4\varepsilon^+_\mathbf{k}}\right)\left(\ln\left|\frac{\varepsilon^2-\varepsilon_c^2}{\varepsilon^2}\right| + i\pi\right),\\
			&\frac{1}{\tau^+_\mathbf{k}} =-\frac{2}{\hbar}\text{Im}\Sigma^+_\mathbf{k}=\frac{n_i u_0^2L^2}{2v_D^2\hbar^3}\left(\varepsilon^+_\mathbf{k} +\frac{\Delta^2}{4\varepsilon^+_\mathbf{k}}\right),
		\end{aligned}
	\end{equation}
	where we assume $u_i$ as a constant $u_0$, $L$ stands for unit length, and $\varepsilon_c$ is the cut-off energy.
	Thus, the dressed vertex is derived as (see Appendix \ref{AppendixB3})
	\begin{equation}
		\tilde{\mathcal{A}} _{\mu ,\mathbf{k}}^{++}(\omega)=\frac{1-B+\frac{\omega}{\hbar}^2\tau^{+2}_{\mathbf{k}}+i\frac{\omega}{\hbar}\tau^{+}_{\mathbf{k}} B}{(1-B)^2+\frac{\omega}{\hbar}^2\tau^{+2}_{\mathbf{k}}} \mathcal{A}^{++}_{\mu,\mathbf{k}},
	\end{equation}
	where $B=\frac{n_i u_0^2L^2}{4\hbar^3 v_D^2}\varepsilon^+_{\mathbf{k}} \sin^2\theta \tau^+_{\mathbf{k}}$.

	%%%%%%%%%%%%%%%%%%%%%%%%%%%%%%%%%%%%%%%%%%%%%%%%%%%%%%%%%%%%%%%%%%%%%%%%%%%%%%%%%%%%%%%%%%%%%%%%%%%%%%%%%%%%%%%%%%%%%%%%%%%
	\begin{figure}[tb]
		\includegraphics[width=1\linewidth]{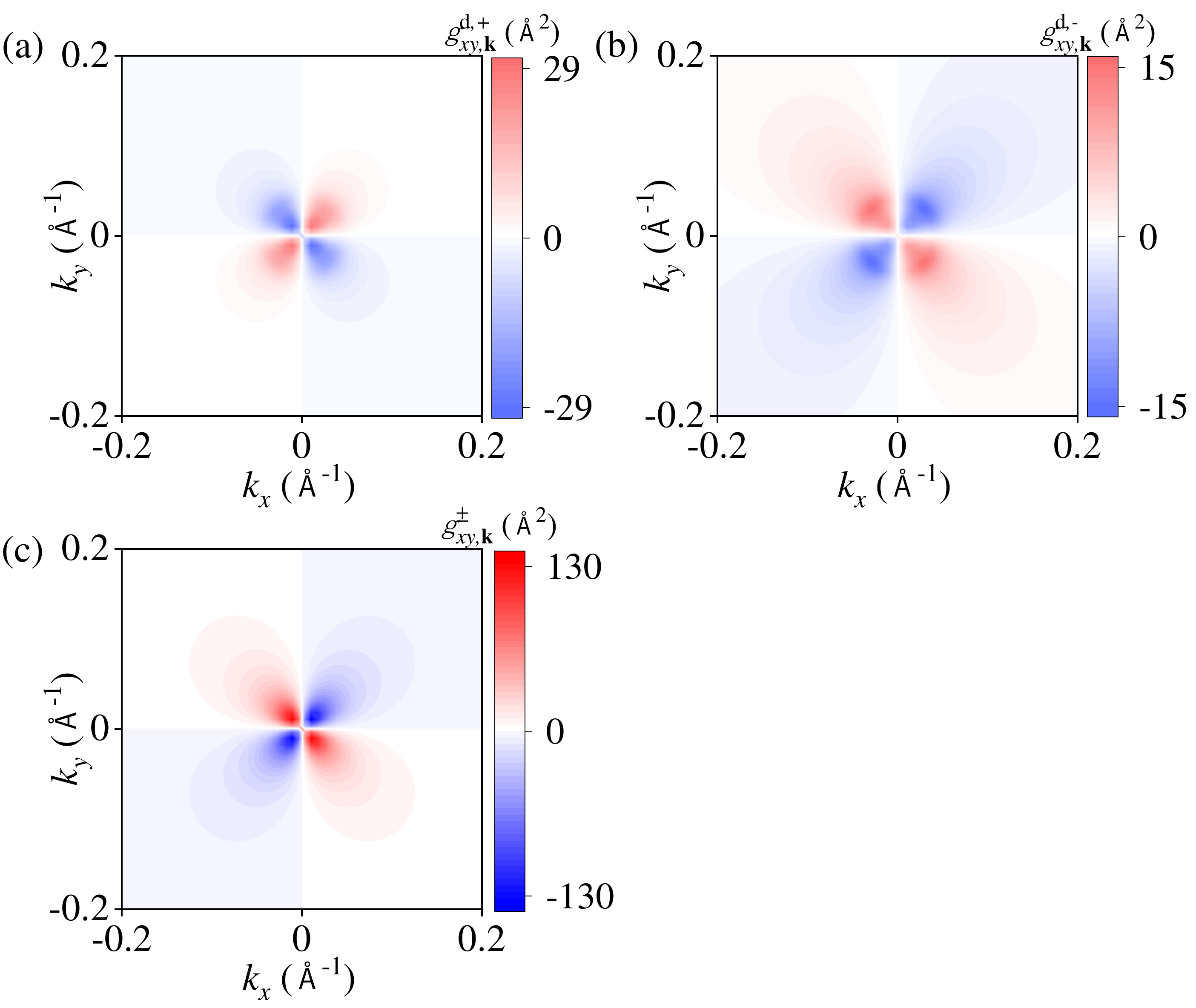}
		\caption{The 2D contour for off-diagonal component of (a), (b) dressed QM $g_{xy,\mathbf{k}}^{\text{d}}$ and (c) QM $ g_{xy,\mathbf{k}}$. (a) is for the $+$ band while (b) is for the $-$ band. The parameters are $\tau = $1, $ \Delta =0.2 $ eV, $n_i u_0^2 L^2 = 10$ $\text{eV}^2 \text{\AA}^2$, $v_D = 10^6$ m/s, $\varepsilon_f = 0.3$ eV}. 
		\label{fig4}
	\end{figure}
	%%%%%%%%%%%%%%%%%%%%%%%%%%%%%%%%%%%%%%%%%%%%%%%%%%%%%%%%%%%%%%%%%%%%%%%%%%%%%%%%%%%%%%%%%%%%%%%%%%%%%%%%%%%%%%%%%%%%%%%%%%%
	
	\subsection{The effect of disorder on BC}
	
	After tedious steps in Appendix \ref{AppendixB}, we obtain the dressed BC at zero temperature limit in Appendix \ref{AppendixB5} and present the results in Fig. \ref{fig4}. It is easy to see that dressed BC and BC have the same symmetry, both being a circular in the $k_x$-$k_y$ plane.
	This tells us that the impurities scattering does not affect the symmetry of geometric quantity BC in the $\mathbf{k}$ space, although scattering potential energy is angle-dependent.
	The difference, however, is that the bare BC (no dressing) of conduction (valence) band [Fig. \ref{fig3}(b),(d)] is negative (positive) as a whole, while both the dressed BC of conduction and valence band [Fig. \ref{fig3}(a),(c)] are positive.
	Therefore, the scattering will decrease the absolute value of conduction band BC but increase the absolute value of valence band BC.
	In addition, there are also some differences in trend along the radial between dressed BC and BC. For conduction (valence) band, BC is minimum (maximum) at $k_x = k_y =0$ and gradually increases (decreases) as $\vert \mathbf{k} \vert$ increases. In contrast, dressed BC first increases sharply from $k_x = k_y =0$ and then decreases gradually with the increase of $\vert \mathbf{k} \vert$.
	Numerically,  when impurity $n_i u_0^2 L^2 = 10$ $\text{eV}^2 \text{\AA}^2$, the dressed BC is two order of magnitude smaller than bare BC.
	%Also dressed BC of conduction band is about three times the valence band. The reason is that the Fermi surface is located in the conduction band, valence band is full. It is difficult to transition to the valence band in the process of scattering.
	For diagonal terms, both bare BC and the dressed BC are zero according to Eq. (\ref{eq7}).

	\begin{figure}[tb]
		\includegraphics[width=1\linewidth]{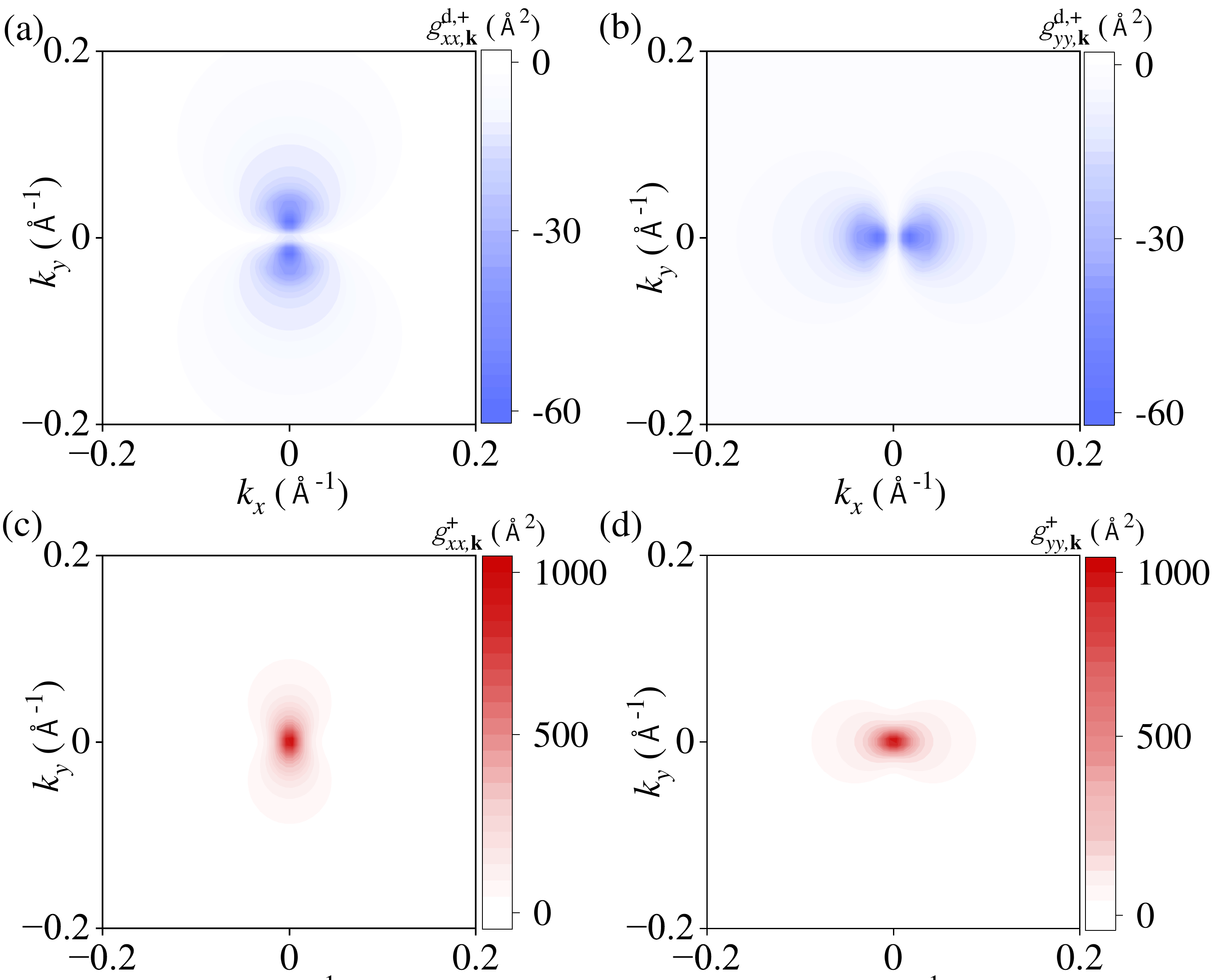}
		\caption{The 2D contour for diagonal components of (a),(b) dressed QM $g_k^{d,+}$ and (c),(d) QM $g_k^+$. (a)(c) are for $xx$ components while (b)(d) are for $yy$ components. The parameters are $\tau = $1, $ \Delta =0.2 $ eV, $n_i u_0^2 L^2 = 10$ $\text{eV}^2 \text{\AA}^2$, $v_D = 10^6$ m/s, $\varepsilon_f = 0.3$ eV}. 
		\label{fig5}
	\end{figure}
	%%%%%%%%%%%%%%%%%%%%%%%%%%%%%%%%%%%%%%%%%%%%%%%%%%%%%%%%%%%%%%%%%%%%%%%%%%%%%%%%%%%%%%%%%%%%%%%%%%%%%%%%%%%%%%%%%%%%%%%%%%%
	
	\subsection{The effect of disorder on QM}
	Fig. \ref{fig4} illustrates the two-dimensional (2D) contour plot for the off-diagonal elements of the dressed QM $g_{xy,\mathbf{k}}^{\text{d},+/-}$ and bare QM $g_{xy,\mathbf{k}}^{\pm}$ (see Appendix \ref{AppendixB5}).
	It is seen both dressed QM and bare QM are of the same symmetry.
	There are in each graph two node lines where bare QM is $0$, centered at $k_x = k_y =0$, and crossing through $\mathbf{k}$ space.
	All of them are symmetric with respect to the vertical mirror along $k_x=k_y$ or $k_x=-k_y$ and anti-symmetric with respect to those along $k_x=0$ or $k_y = 0$.
	Interestingly, while the bare QM and the dressed QM for the valence band have a similar distribution, and both are negative in the first and third quadrants and positive in the second and fourth quadrants; while the dressed QM for the conduction band is of opposite sign.
	It means that the disorder scattering can reduce the absolute value of QM for the conduction band ($+$ band) but increase the absolute value of QM for valence band ($-$ band).
	This result is consistent with BC.
	It is found that the dressed QM is an order of magnitude smaller than the bare QM.
	%And for the same reason as dressed BC, dressed QM of the conduction band is twice as large as valence band.

	The diagonal bare QM and dressed QM for conduction band are shown in Fig. \ref{fig5}. % which show the same symmetry.
	All the four quantities, i.e. $g_{xx/yy,\mathbf{k}}^{d,+}$, $g_{xx/yy,\mathbf{k}}^{+}$ are of two symmetry vertical mirrors along $k_{x}=0$ or $k_{y}=0$.
	The difference is that the nodal lines for $g_{xx,\mathbf{k}}^{d,+}$ and $g_{yy,\mathbf{k}}^{d,+}$ lie along $k_{y}=0$ and $k_{x}=0$, respectively.
	Another feature is that there are a two-fold axis along $k_{y}$ and $k_{x}$, respectively, for $g_{xx,\mathbf{k}}^{d,+}$ (and $g_{xx,\mathbf{k}}^{+}$ ) and $g_{yy,\mathbf{k}}^{d,+}$ (and $g_{yy,\mathbf{k}}^{+}$ ).
	%
	%for intrinsic diagonal QM. They only have two axes of symmetry along $k_x=0$ and $k_y =0$.
	%
	Also, both $g_{xx,\mathbf{k}}^{+}$ and $g_{xx,\mathbf{k}}^{\text{d}, +}$ are extending in the $y$ direction, and both $g_{yy,\mathbf{k}}^{+}$ and $g_{yy,\mathbf{k}}^{\text{d}, +}$ are extending in the $x$ direction. These distributions in momentum space of the $xx$ and $yy$ components of QM look like atomic orbitals of $p_{x}$ and $p_{y}$ and probably of the symmetry of $p_{x}$ and $p_{y}$ in the momentum space.
	In other words, the rotation of $g_{xx,\mathbf{k}}^{+}$ by $\pi /2$ along the $k_z$ axis completely yields $g_{yy,\mathbf{k}}^{+}$, which is also true for dressed QM.
	Nevertheless, the maximum value of QM is located at the center of $k_x = k_y =0$; while dressed QM approaches zero at that point, and the extreme value is located on both side of the original point. %The diagonal terms of dressed QM is an order of magnitude smaller than QM and make it smaller.
	%Also dressed QM is twice as large as QM.

	%Since the effect of valence band is negligible, we put those in the Appendix \ref{AppendixB4} to show. Remarkably, the effect of disorder on QM is very different from the BC except that neither contribute at the origin point in the $k$ space. Generally speaking, the off-diagonal component of dressed QM, $g_{xy,k}^{\text{d},+}$, is about $0.1$ times as many as $g_{xy,k}^+$ while the $g_{xx/yy,k}^{\text{d},+}$ is $100$ times smaller than $g_{xx/yy,k}^{+} $.
	%In terms of symmetry, $g_{xy,k}^{\text{d}, +}$ [Fig. \ref{fig5}(a)] and $g_{xy,k}^{+}$  [Fig. \ref{fig5}(b)] are all symmetry along the $k_x=k_y$ or $k_x=-k_y$ but only $g_{xy,k}^{+}$ is anti-symmetry along the $k_x=0$ or $k_y = 0$. On the other hand, $g_{xy,k}^{\text{d}, +}$  is max or min near the original point while $g_{xy,k}^{\text{d},+}$ rapidly increase from 0 to maximum ($ \approx 0.17 \text{\AA}$) and then gradually decrease as the $k$ increases.
	%Next for the diagonal terms, $ g_{xx/yy,k}^{+}$ is symmetric along either $k_x=0$ or $k_y =0$ [Fig. \ref{fig5}(d) and (f)]while $ g_{xx/yy,k}^{\text{d},+}$ is symmetry along the dash gray line in the  Fig. \ref{fig5}(c) and (e).

	\section{CONCLUSION AND DISCUSION}
	\label{sec5}
	With the help of FBA and dressed vertices, we calculate the eight Feynman diagrams to show how disorder side-jump scattering affects BC and QM. In the gapped graphene, we find that disorder does not affect the symmetry in $\mathbf{k}$ space but affect the  magnitude of BC and QM. Furthermore, impurities may affect some phenomena, which are mainly induced by the BC or QM, such as anomalous Hall effect \cite{Sinitsyn2007}, nonlinear Hall effect \cite{Du2021}, etc. Our calculations provide an idea to explain how impurities affect transport, which is expected to be further understood in the future.

	In the process of calculation, we only consider all first-order diagrams describing side jump scattering, where the effect of disorder on BC and QM is about $0.1$ times as many as those with non-interaction. As for skew scattering for second-order diagrams, proportional to $n_i u_1^3$ or $(n_iu_0^2)^2$ \cite{Sinitsyn2007}, which is not mentioned in this work. Hopefully we can discuss more about how they affect these geometric quantities.
	On the other hand, there are also other interactions, such as electron-electron \cite{Kotov2012} or electron-phonon \cite{Park2008} interactions, can be studied according to Eqs. (\ref{eq4})-(\ref{eq7}). Moreover,  herein we just applied this theory to a two-band graphene, a linear approximation Dirac model. In the future, we will consider the effect of impurities scattering in the twist graphene \cite{Chen2024,Smith2022}, which might be more conducive to explore the contribution of disorder scattering. Also, we noticed that there is a reference discussing the influence of impurities on topological order \cite{Oliveira2024}, but here we only calculate the effect of impurities on BC and QM, which are not equivalent to topological order.
	
	It is well known one can obtain the effect of disorder on BC by measuring quantum anomalous Hall conductivity \cite{Xiao2010}, which is the integral of BC immediately.
	Although QM is of equal importance, it has been less explored until now.
	Generally speaking, as an intermediate physical quantity of multiple coefficients, including intrinsic nonlinear Hall conductivity \cite{Ortix2021,Du2021}, orbital magnetic susceptibilities \cite{Gao2014}, flat-band superconductivity \cite{Chen2024}, linear displacement current \cite{Xiang2024}, QM is difficult to measure directly due to being coupled with other complex functions.
	But there have been continuous efforts to measure QM directly in different systems \cite{Yu2019,Gianfrate2020}.
	Thus, it is expected that our results may be explained by experiments to prove the effect of disorder on QM.

	\begin{acknowledgments}
		This work is supported in part by the Training Program of Major Research plan of the National Natural Science Foundation of China (Grant No. 92165105), and CAS Project for Young Scientists in Basic ResearchGrant No. YSBR-057, the NSFC (Grant No. 11974348 and No. 11834014), and the Strategic Priority Research Program of CAS (Grant No. XDB28000000 and No. XDB33000000).
	\end{acknowledgments}

	%\clearpage
	\appendix
	\setcounter{figure}{0}
	\renewcommand{\thefigure}{A\arabic{figure}}
	\renewcommand{\thefigure}{B\arabic{figure}}
	\begin{widetext}
		\section{The summation on Matsubara frequencies in Feynman diagrams}
		\label{AppendixA}
		
		For the diagram (iii)  in Fig. \ref{fig1}, we get
		
		\begin{equation}
			S_{\mathbf{k},\text{iii}}(\omega )=\sum_{\mathbf{k}^{\prime}}{{\tilde{\mathcal{A}}_{x,\mathbf{k}}^{++}\left( \omega \right)}}\mathcal{A} _{y,\mathbf{k}^{\prime}}^{+-}{\langle V_{\mathbf{k}\mathbf{k}'}^{++}V_{\mathbf{k}'\mathbf{k}}^{-+}\rangle_{\text{imp}} }\frac{1}{\beta}\sum_{ip}^{}{G_{\mathbf{k}}^{R,+}\left( ip+i\omega \right) G_{\mathbf{k}^{\prime}}^{R,+}\left( ip+i\omega \right) G_{\mathbf{k}^{\prime}}^{A,-}\left( ip \right) G_{\mathbf{k}}^{A,+}\left( ip \right)},
		\end{equation}
		where 	$S_{\mathbf{k},\text{iii}}(\omega )$ represents Feynman diagram (iii) in Fig. \ref{fig1}, $\beta=1/k_BT$,  $V_{\mathbf{k}\mathbf{k'}}^{nn'}=\langle n,\mathbf{k}|V(\mathbf{r})|n',\mathbf{k'} \rangle $ ($n,n'=\pm$) is the scattering matrix element of single impurity potential, $  G_{\mathbf{k}}^{R/A} $is the retarded (advanced)  Green's function.
		The summation of Matsubara frequencies of Green's function as
		\begin{equation}
			\begin{aligned}
				I_{\mathbf{k}\mathbf{k}^{\prime},\text{iii}}(\omega ) =&\frac{1}{\beta}\sum_{ip}^{}{G_{\mathbf{k}}^{R,+}\left( ip+i\omega \right) G_{\mathbf{k}^{\prime}}^{R,+}\left( ip+i\omega \right) G_{\mathbf{k}^{\prime}}^{A,-}\left( ip \right) G_{\mathbf{k}}^{A,+}\left( ip \right)}
				\\
				=& \frac{1}{\beta}\sum_{ip}{\frac{1}{ip+i\omega -\varepsilon _{\mathbf{k}}^{+}}}\frac{1}{ip+i\omega -\varepsilon _{\mathbf{k}^{\prime}}^{+}}\frac{1}{ip-\varepsilon _{\mathbf{k}^{\prime}}^{-}}\frac{1}{ip-\varepsilon _{\mathbf{k}}^{+}}
				\\
				=&f(\varepsilon _{\mathbf{k}}^{+})\frac{1}{\varepsilon _{\mathbf{k}}^{+}-\varepsilon _{\mathbf{k}^{\prime}}^{+}}\frac{1}{i\omega +\varepsilon _{\mathbf{k}^{\prime}}^{-}-\varepsilon _{\mathbf{k}}^{+}}\frac{1}{i\omega}+f(\varepsilon _{\mathbf{k}^{\prime}}^{+})\frac{1}{\varepsilon _{\mathbf{k}^{\prime}}^{+}-\varepsilon _{\mathbf{k}}^{+}}\frac{1}{i\omega +\varepsilon _{\mathbf{k}^{\prime}}^{-}-\varepsilon _{\mathbf{k}^{\prime}}^{+}}\frac{1}{i\omega +\varepsilon _{\mathbf{k}}^{+}-\varepsilon _{\mathbf{k}^{\prime}}^{+}}
				\\
				&+f(\varepsilon _{\mathbf{k}^{\prime}}^-)\frac{1}{\varepsilon _{\mathbf{k}^{\prime}}^- -\varepsilon _{\mathbf{k}}^{+}}\frac{1}{i\omega +\varepsilon _{\mathbf{k}^{\prime}}^{-}-\varepsilon _{\mathbf{k}}^{+}}\frac{1}{i\omega +\varepsilon _{\mathbf{k}^{\prime}}^{-}-\varepsilon _{\mathbf{k}^{\prime}}^{+}}+f(\varepsilon _{\mathbf{k}}^{+})\frac{1}{\varepsilon _{\mathbf{k}}^{+}-\varepsilon _{\mathbf{k}^{\prime}}^{-}}\frac{1}{i\omega +\varepsilon _{\mathbf{k}}^{+}-\varepsilon _{\mathbf{k}^{\prime}}^{+}}\frac{1}{i\omega}
				\\
				\xrightarrow{i\omega \rightarrow \omega +i\eta} &\frac{f(\varepsilon _{\mathbf{k}}^{+})}{\varepsilon _{\mathbf{k}}^{+}-\varepsilon _{\mathbf{k}^{\prime}}^{+}}\frac{1}{\varepsilon _{\mathbf{k}^{\prime}}^{-}-\varepsilon _{\mathbf{k}}^{+}+\omega +i\eta}\frac{1}{\omega +i\eta}+\frac{f(\varepsilon _{\mathbf{k}^{\prime}}^{+})}{\varepsilon _{\mathbf{k}^{\prime}}^{+}-\varepsilon _{\mathbf{k}}^{+}}\frac{1}{\varepsilon _{\mathbf{k}^{\prime}}^- -\varepsilon _{\mathbf{k}^{\prime}}^{+}+\omega +i\eta}\frac{1}{\varepsilon _{\mathbf{k}}^{+}-\varepsilon _{\mathbf{k}^{\prime}}^{+}+\omega +i\eta}
				\\
				&+\frac{f(\varepsilon _{\mathbf{k}^{\prime}}^{-})}{\varepsilon _{\mathbf{k}^{\prime}}^{-}+\omega +i\eta -\varepsilon _{\mathbf{k}}^{+}}\frac{1}{\varepsilon _{\mathbf{k}^{\prime}}^{-}+\omega +i\eta -\varepsilon _{\mathbf{k}^{\prime}}^{+}}\frac{1}{\varepsilon _{\mathbf{k}^{\prime}}^{-}-\varepsilon _{\mathbf{k}}^{+}}+\frac{f(\varepsilon _{\mathbf{k}}^{+})}{\omega +i\eta}\frac{1}{\varepsilon _{\mathbf{k}}^{+}+\omega +i\eta -\varepsilon _{\mathbf{k}^{\prime}}^{+}}\frac{1}{\varepsilon _{\mathbf{k}}^{+}-\varepsilon _{\mathbf{k}^{\prime}}^{-}}
				\\
				=&\frac{f(\varepsilon _{\mathbf{k}}^{+})}{\varepsilon _{\mathbf{k}}^{+}-\varepsilon _{\mathbf{k}^{\prime}}^{+}}\left[ \frac{1}{\omega +\varepsilon _{\mathbf{k}^{\prime}}^{-}-\varepsilon _{\mathbf{k}}^{+}}-i\pi \delta \left( \omega +\varepsilon _{\mathbf{k}^{\prime}}^{-}-\varepsilon _{\mathbf{k}}^{+} \right) \right] \left[ \frac{1}{\omega}-i\pi \delta \left( \omega \right) \right]
				\\
				&+\frac{f(\varepsilon _{\mathbf{k}^{\prime}}^{+})}{\varepsilon _{\mathbf{k}^{\prime}}^{+}-\varepsilon _{\mathbf{k}}^{+}}\left[ \frac{1}{\omega +\varepsilon _{\mathbf{k}^{\prime}}^{-}-\varepsilon _{\mathbf{k}^{\prime}}^{+}}-i\pi \delta \left( \omega +\varepsilon _{\mathbf{k}^{\prime}}^- -\varepsilon _{\mathbf{k}^{\prime}}^{+} \right) \right] \left[ \frac{1}{\omega +\varepsilon _{\mathbf{k}}^{+}-\varepsilon _{\mathbf{k}^{\prime}}^{+}}-i\pi \delta \left( \omega +\varepsilon _{\mathbf{k}}^{+}-\varepsilon _{\mathbf{k}^{\prime}}^{+} \right) \right]
				\\
				&+\frac{f(\varepsilon _{\mathbf{k}^{\prime}}^{-})}{\varepsilon _{\mathbf{k}^{\prime}}^{-}-\varepsilon _{\mathbf{k}}^{+}}\left[ \frac{1}{\omega +\varepsilon _{\mathbf{k}^{\prime}}^{-}-\varepsilon _{\mathbf{k}}^{+}}-i\pi \delta \left( \omega +\varepsilon _{\mathbf{k}^{\prime}}^- -\varepsilon _{\mathbf{k}}^{+} \right) \right] \left[ \frac{1}{\omega +\varepsilon _{\mathbf{k}^{\prime}}^{-}-\varepsilon _{\mathbf{k}^{\prime}}^{+}}-i\pi \delta \left( \omega +\varepsilon _{\mathbf{k}^{\prime}}^{-}-\varepsilon _{\mathbf{k}^{\prime}}^{+} \right) \right]
				\\
				&+\frac{f(\varepsilon _{\mathbf{k}}^{+})}{\varepsilon _{\mathbf{k}}^{+}-\varepsilon _{\mathbf{k}^{\prime}}^{-}}\left[ \frac{1}{\omega +\varepsilon _{\mathbf{k}}^{+}-\varepsilon _{\mathbf{k}^{\prime}}^{+}}-i \pi \delta \left( \omega +\varepsilon _{\mathbf{k}}^{+}-\varepsilon _{\mathbf{k}^{\prime}}^{+} \right) \right] \left[ \frac{1}{\omega}-i\pi \delta \left( \omega \right) \right].
			\end{aligned}
		\end{equation}
		where we use the analytical continuation of Green's functions $i\omega \rightarrow \omega +i\eta$, where the methods of analytical continuation and the summation of Matsubara frequencies are adopted from Ref.\onlinecite{Bruss2003,Parker2019}.

		\section{Some calculations in two-band gapped graphene}
		\label{AppendixB}
		
		\subsection{Disorder average}
		\label{AppendixB1}
		The Hamiltonian for two-band gapped graphene is
		\begin{equation}
			H=\\
			\begin{aligned}
				\begin{bmatrix}
					\frac{\Delta}{2} & v_D\hbar (\tau k_x -i k_y ) \\
					v_D\hbar (\tau k_x +i k_y ) & -\frac{\Delta}{2}
				\end{bmatrix},
			\end{aligned}
		\end{equation}
		Let $K_1=\tau v_D\hbar k_x$, $K_2=v_D\hbar k_y$, $K_3=\frac{\Delta}{2}$, $K=\sqrt{K_1^2+K_2^2+K_3^2}$, $\tan \varphi=\frac{K_2}{K_1}$, $\cos \theta=\frac{K_3}{K}$, $a=\sqrt{\frac{1+\cos \theta}{2}}$, $b=\sqrt{\frac{1-\cos \theta}{2}}$.
		Thus, the wave function can be written as
		\begin{equation}
			\begin{aligned}
				|+ \rangle &=\frac{1}{\sqrt{2K(K+K_3)}}\left(\begin{array}{c} K+K_3 \\ K_1+iK_2 \end{array}\right) =\left(\begin{array}{c} a \\ be^{i \varphi} \end{array}\right),\\
				|- \rangle &=\frac{1}{\sqrt{2K(K-K_3)}}\left(\begin{array}{c} -K+K_3 \\ K_1+iK_2 \end{array}\right) =\left(\begin{array}{c} -b \\ ae^{i \varphi} \end{array}\right),
			\end{aligned}
		\end{equation}
		where the $+$ and $-$ indicate the upper and lower energy band. The Berry connection for  inter and intraband are
		\begin{equation}
			\begin{aligned}
				A_{x/y,\mathbf{k}}^{+-} &=\left. \langle +\left| i\partial _{x/y} \right|- \right. \rangle =\frac{\tau v_{D}^{2}\hbar ^2k_{y/x}-i\frac{\Delta}{2}v_{D}^{2}\hbar ^2k_{x/y}/K}{2\sin \theta K^2},
				\\
				A_{x/y,\mathbf{k}}^{++}&=\left. \langle +\left| i\partial _{x/y} \right|+ \right. \rangle=\frac{\tau v_{D}^{2}\hbar ^2k_{y/x}}{2K(K+\frac{\Delta}{2})}.
			\end{aligned}
		\end{equation}
		
		The expectation value of impurity scattering in band $|+\rangle$ can be written as \cite{Sinitsyn2007}
		\begin{equation}
			\begin{aligned}
				V_{\mathbf{k}\mathbf{k}^{\prime}}^{++} &=\langle +,\mathbf{k}|V(\mathbf{r})|+,\mathbf{k}^{\prime}\rangle
				=\int{d}r\left( a,be^{-i\varphi} \right) e^{-i\mathbf{k}\mathbf{r}}\sigma _0\Sigma _iu_i\delta \left( \mathbf{r}-\mathbf{R}_i \right) \left( \begin{array}{c}
					a^{\prime}\\
					b^{\prime}e^{i\varphi ^{\prime}}\\
				\end{array} \right) e^{i\mathbf{k}^{\prime}\mathbf{r}}		\\
				&=\Sigma _iu_i\left[\int dr e^{-i\mathbf{k}\mathbf{r}} e^{i\mathbf{k}^{\prime}\mathbf{r}}\delta \left( \mathbf{r}-\mathbf{R}_i \right)\right]\left( a,be^{-i\varphi} \right)
				\sigma _0\left( \begin{array}{c}
					a^{\prime}\\
					b^{\prime}e^{i\varphi ^{\prime}}\\
				\end{array} \right)
				=\Sigma _iu_ie^{i\left( \mathbf{k}^{\prime}-\mathbf{k} \right) \mathbf{R}_i}\left[ aa^{\prime}+bb^{\prime}e^{i\left( \varphi ^{\prime}-\varphi \right)} \right],
			\end{aligned}
			\label{eqB4}
		\end{equation}
		other components such as $V_{\mathbf{k}\mathbf{k}^{\prime}}^{+-}$, $V_{\mathbf{k}\mathbf{k}^{\prime}}^{+-}$ and $V_{\mathbf{k}\mathbf{k}^{\prime}}^{-+}$ can be calculated in a similar way and are not shown explicitly. 
		And
		\begin{equation}
			\begin{aligned}
				\langle V_{\mathbf{k}\mathbf{k}^{\prime}}^{++}V_{\mathbf{k}^{\prime}\mathbf{k}}^{++}\rangle_\text{imp} = n_iu_{0}^{2}\left[ a^2a^{\prime2}+b^2b^{\prime2}+2aa^{\prime}bb^{\prime}\cos \left( \varphi ^{\prime}-\varphi \right) \right],
			\end{aligned}
		\end{equation}
		where $a', b', \varphi^\prime$ are all dependent on the $\mathbf{k}^\prime$.
		
		\subsection{Self-energy and relaxation time}
		\label{AppendixB2}
		In eigenstate representation, following the Fourier transform of Eq. (\ref{eqB4}), the impurity scattering matrix can be written as
		\begin{equation}
			\begin{aligned}
				&\langle \varphi,\mathbf{k}|V(\mathbf{r})|\varphi,\mathbf{k}' \rangle=
				\left(\begin{array}{ll}
					V^{++} & V^{+-} \\
					V^{-+} & V^{--}
				\end{array}\right)
				=\Sigma _iu_ie^{i\left( \mathbf{k}^{\prime}-\mathbf{k} \right) \mathbf{R}_i}
				\left(\begin{array}{ll}
					aa'+bb'e^{i(\varphi'-\varphi)} & -ab'+a'be^{i(\varphi'-\varphi)} \\
					-a'b+ab'e^{i(\varphi'-\varphi)} & bb'+aa'e^{i(\varphi'-\varphi)}
				\end{array}\right).
			\end{aligned}
		\end{equation}
		Also Green's function become diagonal with respect to the band index in eigenstate representation,
		\begin{equation}
			\langle\varphi,\mathbf{k}'|G_\mathbf{k}|\varphi,\mathbf{k}'\rangle =
			\left(\begin{array}{ll}
				G_\mathbf{k}^+ & 0 \\
				0 & G_\mathbf{k}^-
			\end{array}\right)=
			\left(\begin{array}{ll}
				\frac{1}{ip -\varepsilon_\mathbf{k}} & 0 \\
				0 & \frac{1}{ip+\varepsilon_\mathbf{k}}
			\end{array}\right).
		\end{equation}
		Because the translation invariant is recovered after the disorder average, $\Sigma$ become diagonal with respect to the wave vector.
		Thus \begin{equation}
			\begin{aligned}
				\Sigma
				&=\langle VGV\rangle_{\text{imp}}  =\left(\begin{array}{ll}
					\Sigma^+ & 0 \\
					0 & \Sigma^-
				\end{array}\right) \\
			\end{aligned},
		\end{equation}
		where
		\begin{equation}
			\begin{aligned}
				\Sigma ^{\pm} &=\langle V_{\mathbf{k} \mathbf{k}'}^{\pm +}V_{\mathbf{k}' \mathbf{k}}^{+\pm}\rangle _\text{imp}G_\mathbf{k}^{+}+\langle V_{\mathbf{k}\mathbf{k}'}^{\pm -}V_{\mathbf{k}' \mathbf{k}}^{-\pm}\rangle _\text{imp}G_\mathbf{k}^{-}
				=-\frac{n_iu_{0}^{2}L^2}{4\pi v_{D}^{2}\hbar ^2}(\varepsilon_\mathbf{k}^\pm  \pm \frac{\Delta ^2}{4\varepsilon_\mathbf{k}})
				\ln(\frac{\varepsilon_\mathbf{k}^{\pm 2}-\varepsilon _{c}^{2}}{\varepsilon_\mathbf{k}^{\pm 2}}).
			\end{aligned}
		\end{equation}
		Then we obtain in the first Born approximation
		\begin{equation}
			\begin{aligned}
				&\Sigma^{ \pm}=-\frac{n_i u_0^2L^2}{4\pi v_D^2\hbar^2}(\varepsilon_\mathbf{k}^\pm \pm \frac{\Delta^2}{4\varepsilon_\mathbf{k}^\pm })(\ln\left|\frac{\varepsilon_\mathbf{k}^{\pm 2}-\varepsilon_c^2}{\varepsilon^{\pm 2}}\right| + i\pi),\\
				&\frac{1}{\tau^+_\mathbf{k}} =-\frac{2}{\hbar}\text{Im}\Sigma^+_\mathbf{k}=\frac{n_i u_0^2L^2}{2\hbar v_D^2\hbar^2}(\varepsilon^+_\mathbf{k} +\frac{\Delta^2}{4\varepsilon^+_\mathbf{k}}).\\
			\end{aligned}
		\end{equation}

		\subsection{The corrected vertices}
		\label{AppendixB3}
		Take the $\tilde{\mathcal{A}}_{x,\mathbf{k}}^{+-}(\omega )$ as an example.
		In the weak disorder limit and $\varepsilon \ll \hbar\omega $ we have \cite{Du2021}
		\begin{equation}
			\mathscr{G}^R_\mathbf{k}(\varepsilon+\hbar\omega)\mathscr{G}^A_\mathbf{k}(\varepsilon) \approx  \frac{2\pi}{\hbar}\frac{\tau_\mathbf{k}}{1-i\frac{\omega}{\hbar}\tau_\mathbf{k}}\delta(\varepsilon-\varepsilon_\mathbf{k}),
		\end{equation}
		then,
		\begin{equation}
			\begin{aligned}
				\tilde{\mathcal{A}} _{x,\mathbf{k}}^{+-}(\omega ) &=\mathcal{A} _{\mu, \mathbf{k}}^{++}
				+\sum_{\mathbf{k}''}{\langle }V_{\mathbf{k}''\mathbf{k}}^{++}V_{\mathbf{k}\mathbf{k}''}^{++}\rangle_{\text{imp}} \mathscr{G} _{\mathbf{k}''}^{R,+}\mathscr{G} _{\mathbf{k}''}^{A,+}\tilde{\mathcal{A}}  _{x,\mathbf{k}''}^{+-}\left( \omega \right)
				\\
				&\approx \mathcal{A} _{x,\mathbf{k}}^{++}+\frac{n_iu_{0}^{2}L^2}{(2\pi )^2}\int{\mathbf{k}}''d\mathbf{k}''\int{d}\varphi ''[a^2a''^2+b^2b''^2+2aa''bb''\cos(\varphi ''-\varphi )]\frac{2\pi}{\hbar}\frac{\tau _{\mathbf{k}''}^{+}}{1-i\frac{\omega}{\hbar} \tau _{\mathbf{k}''}^{+}}\delta (\varepsilon -\varepsilon _{\mathbf{k}''}^{+})\tilde{\mathcal{A}}_{x,\mathbf{k}''}^{+-}\left( \omega \right)
				\\
				&=\frac{\tau v_D\hbar \sin\theta \sin\varphi}{2(\varepsilon _{\mathbf{k}}^{+}+\frac{\Delta}{2})}+\frac{n_iu_{0}^{2}L^2}{2\hbar}\int{\mathbf{k}}''d\mathbf{k}''2aa''bb''\frac{\tau _{\mathbf{k}''}^{+}}{1-i\frac{\omega}{\hbar}\tau _{\mathbf{k}''}^{+}}\delta (\varepsilon _{\mathbf{k}}^{+}-\varepsilon _{\mathbf{k}''}^{+})\tilde{\mathcal{A}}  _{x,\mathbf{k}''}^{+-}\left( \omega \right)
				\\
				&=\frac{\tau v_D\hbar \sin\theta \sin\varphi}{2(\varepsilon _{\mathbf{k}}^{+}+\frac{\Delta}{2})}+\frac{n_iu_{0}^{2}L^2}{4\hbar v_{D}^{2}\hbar ^2}\varepsilon _{\mathbf{k}}^{+}\sin^2\theta \frac{\tau _{\mathbf{k}}^{+}}{1-i\frac{\omega}{\hbar} \tau _{\mathbf{k}}^{+}}\tilde{\mathcal{A}}_{x,\mathbf{k}''}^{+-}\left( \omega \right).
			\end{aligned}
		\end{equation}
		And
		\begin{equation}
			\begin{aligned}
				\tilde{\mathcal{A}}_{x,\mathbf{k}}^{++}(\omega ) =\frac{1-B+\frac{\omega}{\hbar}^2\tau^{+2}_{\mathbf{k}}+i\frac{\omega}{\hbar}\tau^{+}_{\mathbf{k}} B}{(1-B)^2
					+\frac{\omega}{\hbar}^2\tau^{+2}_{\mathbf{k}}} \mathcal{A}^{++}_{x,\mathbf{k}},\\
				\tilde{\mathcal{A}} _{y,\mathbf{k}}^{++}(\omega ) =\frac{1-B+\frac{\omega}{\hbar}^2\tau^{+2}_{\mathbf{k}}+i\frac{\omega}{\hbar}\tau^{+}_{\mathbf{k}} B}{(1-B)^2+\frac{\omega}{\hbar}^2\tau^{+2}_{\mathbf{k}}} \mathcal{A}^{++}_{y,\mathbf{k}},
			\end{aligned}
		\end{equation}
		where  $B=\frac{n_i u_0^2L^2}{4\hbar^3 v_D^2 }\varepsilon^+_{\mathbf{k}} \sin^2\theta \tau^+_{\mathbf{k}}$ and $L= 1 \text{\AA}$.

		\subsection{The calculations of dressed BC and QM}
		\label{AppendixB4}
		According to Eq. (\ref{eq8}), for the  diagram (iii) we can get
		\begin{equation}
			\chi^+_{xy ,\mathbf{k},\text{iii}}=\int_0^{\infty}S_{\mathbf{k},\text{iii}}(\omega )d\omega .
		\end{equation}
		Considering the reciprocity of integration and summation, we deal with the integral part first,
		\begin{equation}
			\chi ^+_{xy ,\mathbf{\mathbf{k}},\text{iii}}=\sum_{\mathbf{\mathbf{k'}} }{\mathcal{A} _{y,\mathbf{\mathbf{k}}^{\prime}}^{+-}}\langle  V_{\mathbf{k}\mathbf{k}^{\prime}}^{++}V_{\mathbf{k}^{\prime} \mathbf{k}}^{-+}\rangle_{\text{imp}} \int_0^{\infty}{\tilde{\mathcal{A}}_{x,\mathbf{k}}^{++}\left( \omega \right) I_{\mathbf{k}\mathbf{k}^{\prime},\text{iii}}\left( \omega \right) d\omega},
		\end{equation}
		where
		\begin{equation}
			\begin{aligned}
				&\int_0^{\infty}{\tilde{\mathcal{A}}_{x,\mathbf{k}}^{++}\left( \omega \right) I_{\mathbf{k}\mathbf{k}^{\prime},\text{iii}}\left( \omega \right) d\omega}  \\ &=\frac{f(\varepsilon _{\mathbf{k}}^{+})}{\varepsilon _{\mathbf{k}}^{+}-\varepsilon _{\mathbf{k}^{\prime}}^{+}}\int_0^{\infty}{\tilde{\mathcal{A}}_{x,\mathbf{k}}^{++}\left( \omega \right)}\left[ \frac{-i\pi \delta \left( \omega \right)}{\omega +\varepsilon _{\mathbf{k}^{\prime}}^{-}-\varepsilon _{\mathbf{k}}^{+}}+\frac{-i\pi \delta \left( \omega +\varepsilon _{\mathbf{k}^{\prime}}^{-}-\varepsilon _{\mathbf{k}}^{+} \right)}{\omega}-\pi ^2\delta \left( \omega +\varepsilon _{\mathbf{k}^{\prime}}^{-}-\varepsilon _{\mathbf{k}}^{+} \right) \delta \left( \omega \right) \right]
				\\
				&+\frac{f(\varepsilon _{\mathbf{k}^{\prime}}^{+})}{\varepsilon _{\mathbf{k}^{\prime}}^{+}-\varepsilon _{\mathbf{k}}^{+}}\int_0^{\infty}{\tilde{\mathcal{A}}_{x,\mathbf{k}}^{++}\left( \omega \right) \left[ \frac{-i\pi \delta \left( \omega +\varepsilon _{\mathbf{k}}^{+}-\varepsilon _{\mathbf{k}^{\prime}}^{+} \right)}{\omega +\varepsilon _{\mathbf{k}^{\prime}}^{-}-\varepsilon _{\mathbf{k}^{\prime}}^{+}}+\frac{-i\pi \delta \left( \omega +\varepsilon _{\mathbf{k}^{\prime}}^- -\varepsilon _{\mathbf{k}^{\prime}}^{+} \right)}{\omega +\varepsilon _{\mathbf{k}}^{+}-\varepsilon _{\mathbf{k}^{\prime}}^{+}}-\pi ^2\delta \left( \omega +\varepsilon _{\mathbf{k}^{\prime}}^- -\varepsilon _{\mathbf{k}^{\prime}}^{+} \right) \delta \left( \omega +\varepsilon _{\mathbf{k}}^{+}-\varepsilon _{\mathbf{k}^{\prime}}^{+} \right) \right]}
				\\
				&+\frac{f(\varepsilon _{\mathbf{k}^{\prime}}^{-})}{\varepsilon _{\mathbf{k}^{\prime}}^{-}-\varepsilon _{\mathbf{k}}^{+}}\int_0^{\infty}{\tilde{\mathcal{A}}_{x,\mathbf{k}}^{++}\left( \omega \right) \left[ \frac{-i\pi \delta \left( \omega +\varepsilon _{\mathbf{k}^{\prime}}^{-}-\varepsilon _{\mathbf{k}^{\prime}}^{+} \right)}{\omega +\varepsilon _{\mathbf{k}^{\prime}}^{-}-\varepsilon _{\mathbf{k}}^{+}}+\frac{-i\pi \delta \left( \omega +\varepsilon _{\mathbf{k}^{\prime}}^- -\varepsilon _{\mathbf{k}}^{+} \right)}{\omega +\varepsilon _{\mathbf{k}^{\prime}}^{-}-\varepsilon _{\mathbf{k}^{\prime}}^{+}}-\pi ^2\delta \left( \omega +\varepsilon _{\mathbf{k}^{\prime}}^--\varepsilon _{\mathbf{k}}^{+} \right) \delta \left( \omega +\varepsilon _{\mathbf{k}^{\prime}}^{-}-\varepsilon _{\mathbf{k}^{\prime}}^{+} \right) \right]}
				\\
				&+\frac{f(\varepsilon _{\mathbf{k}}^{+})}{\varepsilon _{\mathbf{k}}^{+}-\varepsilon _{\mathbf{k}'}^{-}}\int_0^{\infty}{\tilde{\mathcal{A}}_{x,\mathbf{k}}^{++}\left( \omega \right) \left[ \frac{-i\pi \delta \left( \omega \right)}{\omega +\varepsilon _{\mathbf{k}}^{+}-\varepsilon _{\mathbf{k}^{\prime}}^{+}}+\frac{-i\pi \delta \left( \omega +\varepsilon _{\mathbf{k}}^{+}-\varepsilon _{\mathbf{k}^{\prime}}^{+} \right)}{\omega}-\pi ^2\delta \left( \omega +\varepsilon _{\mathbf{k}}^{+}-\varepsilon _{\mathbf{k}^{\prime}}^{+} \right) \delta \left( \omega \right) \right]},
				\\
				&=-i\pi \tilde{\mathcal{A}}_{x,\mathbf{k}}^{++}\left( \omega  \right) \left[ \frac{f(\varepsilon _{\mathbf{k}}^{+})-f(\varepsilon _{\mathbf{k}'}^{-})}{\left( \varepsilon _{\mathbf{k}}^{+}-\varepsilon _{\mathbf{k}^{\prime}}^{+} \right) \left( \varepsilon _{\mathbf{k}}^{+}-\varepsilon _{\mathbf{k}'}^{-} \right)} \right] \bigg|_{\omega =\varepsilon _{\mathbf{k}}^{+}-\varepsilon _{\mathbf{k}'}^{-}>0}^{}
				-i\pi \tilde{\mathcal{A}}_{x,\mathbf{k}}^{++}\left( \omega  \right) \left[ \frac{f(\varepsilon_{\mathbf{k^\prime}}^{+})-f(\varepsilon _{\mathbf{k}}^{+})}{\left( \varepsilon _{\mathbf{k}}^{+}-\varepsilon _{\mathbf{k}^{\prime}}^{+} \right) \left( \varepsilon _{\mathbf{k}}^{+}-\varepsilon _{\mathbf{k}'}^{-} \right)} \right] \bigg|_{\omega =\varepsilon _{\mathbf{k}'}^{+}-\varepsilon _{\mathbf{k}}^{+}>0}^{}
				\\
				&-i\pi \tilde{\mathcal{A}}_{x,\mathbf{k}}^{++}\left( \omega \right) \left[ \frac{f(\varepsilon _{\mathbf{k'}}^{-})-f(\varepsilon _{\mathbf{k'}}^{+})}{\left( \varepsilon _{\mathbf{k}}^{+}-\varepsilon _{\mathbf{k}^{\prime}}^{+} \right) \left( \varepsilon _{\mathbf{k}}^{+}-\varepsilon _{\mathbf{k}'}^{-} \right)} \right] \bigg|_{\omega =\varepsilon_{\mathbf{k'}}^{+}-\varepsilon_{\mathbf{k}'}^{-}>0}
				-\pi ^2\left[ f\left( \varepsilon _{\mathbf{k}}^{+} \right) \tilde{\mathcal{A}}_{x,\mathbf{k}}^{++}\left( \omega \right) \bigg|_{\omega =0}-f\left( \varepsilon _{\mathbf{k}'}^{-} \right) \tilde{\mathcal{A}}_{x,\mathbf{k}}^{++}\left( \omega \right) \bigg|_{\omega =\varepsilon _{\mathbf{k}}^{+}-\varepsilon _{\mathbf{k}'}^{-}}^{} \right]
				\\
				&\times \frac{\delta \left( \varepsilon _{\mathbf{k}}^{+}-\varepsilon _{\mathbf{k}'}^{+} \right)}{\varepsilon _{\mathbf{k}}^{+}-\varepsilon _{\mathbf{k}'}^{-}}.
			\end{aligned}
		\end{equation}
		Next, we plan to calculate the susceptibility for the diagram (iii) in Fig. \ref{fig1},
		\begin{equation}
			\begin{aligned}
				\chi^+ _{xy ,\mathbf{k},\text{iii}}= &\sum_{\mathbf{k}^{\prime}}{\mathcal{A} _{y,\mathbf{k}^{\prime}}^{+-}}\langle  V_{\mathbf{k}\mathbf{k}^{\prime}}^{++}V_{\mathbf{k}^{\prime} \mathbf{k}}^{-+}\rangle_{\text{imp}} \int_0^{\infty}{\tilde{\mathcal{A}}_{x,\mathbf{k}}^{++}\left( \omega \right) I_{\mathbf{k}\mathbf{k}^{\prime},1}\left( \omega \right) d\omega}
				\\
				=&\frac{L^2}{(2\pi )^2}\int{\mathbf{k}^{\prime}} d \mathbf{k}^{\prime}\int{d}\varphi^{\prime}\frac{-v_D\hbar \varepsilon _{\mathbf{k}^{\prime}}^{+}\cos\varphi^{\prime}-i\frac{\Delta}{2}v_D\sin\varphi^{\prime}}{2\varepsilon _{\mathbf{k}^{\prime}}^{+2}}n_iu_{0}^{2}[a^{\prime} b^2b^{\prime}-a^2a^{\prime} b^{\prime}+aa^{\prime2}be^{i(\varphi -\varphi^{\prime})}-abb^{\prime2}e^{i(\varphi^{\prime}-\varphi )}]
				\\
				&\Bigg\{-i\pi \tilde{\mathcal{A}}_{x,\mathbf{k}}^{++}\left( \omega \right) \left[ \frac{f(\varepsilon _{\mathbf{k}}^{+})-f(\varepsilon _{\mathbf{k}^{\prime}}^{-})}{\left( \varepsilon _{\mathbf{k}}^{+}-\varepsilon _{\mathbf{k}^{\prime}}^{+} \right) \left( \varepsilon _{\mathbf{k}}^{+}-\varepsilon _{\mathbf{k}^{\prime}}^{-} \right)} \right] \bigg| _{\omega =\varepsilon _{\mathbf{k}}^{+}-\varepsilon _{\mathbf{k}^{\prime}}^{-}>0}^{}
				-i\pi \tilde{\mathcal{A}}_{x,\mathbf{k}}^{++}\left( \omega \right) \left[ \frac{f(\varepsilon _{\mathbf{k'}}^{+})-f(\varepsilon _{\mathbf{k}}^{+})}{\left( \varepsilon _{\mathbf{k}}^{+}-\varepsilon _{\mathbf{k}^{\prime}}^{+} \right) \left( \varepsilon _{\mathbf{k}}^{+}-\varepsilon _{\mathbf{k}^{\prime}}^{-} \right)} \right] \bigg| _{\omega =\varepsilon _{\mathbf{k}^{\prime}}^{+}-\varepsilon _{\mathbf{k}}^{+}>0}^{}
				\\
				&-i\pi \tilde{\mathcal{A}}_{x,\mathbf{k}}^{++}\left( \omega \right) \left[ \frac{f(\varepsilon _{\mathbf{k'}}^{-})-f(\varepsilon _{\mathbf{k'}}^{+})}{\left( \varepsilon _{\mathbf{k}}^{+}-\varepsilon _{\mathbf{k}^{\prime}}^{+} \right) \left( \varepsilon _{\mathbf{k}}^{+}-\varepsilon _{\mathbf{k}^{\prime}}^{-} \right)} \right] \bigg| _{\omega =\varepsilon _{\mathbf{k'}}^{+}-\varepsilon _{\mathbf{k}^{\prime}}^{-}>0}\\
				&-\pi ^2\left[ f\left( \varepsilon _{\mathbf{k}}^{+} \right) \tilde{\mathcal{A}}_{x,\mathbf{k}}^{++}\left( \omega \right) \bigg| _{\omega =0}-f\left( \varepsilon _{\mathbf{k}^{\prime}}^{-} \right) \tilde{\mathcal{A}}_{x,\mathbf{k}}^{++}\left( \omega \right) \bigg| _{\omega =\varepsilon _{\mathbf{k}}^{+}-\varepsilon _{\mathbf{k}^{\prime}}^{-}}^{} \right] \frac{\delta \left( \varepsilon _{\mathbf{k}}^{+}-\varepsilon _{\mathbf{k}^{\prime}}^{+} \right)}{\varepsilon _{\mathbf{k}}^{+}-\varepsilon _{\mathbf{k}^{\prime}}^{-}} \Bigg\}.
			\end{aligned}
		\end{equation}
		As we know, at zero temperature limit we have
		\begin{equation}
			f(\varepsilon_\mathbf{k}^+)-f(\varepsilon_{\mathbf{k}'}^+) \approx f'(\varepsilon_{\mathbf{k}'}^+)(\varepsilon_\mathbf{k}^+ -\varepsilon_{\mathbf{k}'}^+) \approx -\delta(\varepsilon_\mathbf{k}^+-\varepsilon_f)(\varepsilon_\mathbf{k}^+-\varepsilon_{\mathbf{k}'}^+).
		\end{equation}
		Thus, we obtain
		\begin{equation}
			\begin{aligned}
				\chi^+_{xy,\mathbf{k},\text{iii}}= &-\frac{iabn_iu_{0}^{2}L^2}{8\hbar v_{D}}\frac{-\varepsilon _{f}\Delta \cos\varphi -i\left[ \varepsilon_{f}^{2}+(\frac{\Delta}{2})^2 \right] \sin\varphi }{\varepsilon_{f}^{2}}\frac{\widetilde{\mathcal{A} }_{x,\mathbf{k}}^{++}\left( \omega \right)}{\varepsilon _{\mathbf{k}}^{+}+\varepsilon _f}\bigg|_{\omega =\varepsilon _f-\varepsilon _{\mathbf{k}}^{+}>0}\\
				&	-\frac{\pi abn_iiu_{0}^{2}L^2}{16\hbar v_{D}}\frac{-\varepsilon _{\mathbf{k}}^{+}\Delta \cos\varphi -i\left[ \varepsilon_{f}^{2}+(\frac{\Delta}{2})^2 \right] \sin\varphi}{\varepsilon _{\mathbf{k}}^{+3}}\left[\widetilde{\mathcal{A} }_{x,\mathbf{k}}^{++}\left( \omega \right) f(\varepsilon _{\mathbf{k}}^{+})\bigg|_{\omega =0}^{}-\widetilde{\mathcal{A} }_{x,\mathbf{k}}^{++}\left( \omega \right) f(-\varepsilon _{\mathbf{k}}^{+})\bigg|_{\omega =\varepsilon _{\mathbf{k}}^{+}+\varepsilon _f>0}^{}\right].
			\end{aligned}
		\end{equation}
		Other three diagrams for $-$ band are
		\begin{equation}
			\begin{aligned}
				\chi^+_{xy,\mathbf{k},\text{iv}}= &-\frac{i\tau abn_iu_{0}^{2}L^2}{8\hbar v_{D}^{}}\frac{\left( \varepsilon_{f}^{2}+\frac{\Delta ^2}{4} \right) i\cos\varphi +\varepsilon _{f}\Delta \sin\varphi }{\varepsilon_{f}^{2}}\frac{\widetilde{\mathcal{A} }_{y,\mathbf{k}}^{++}\left( \omega \right)}{\varepsilon _\mathbf{k}^+ +\varepsilon _f}\bigg|_{\omega =\varepsilon _f-\varepsilon _{\mathbf{k}}^{+}>0}
				\\
				&-\frac{\pi \tau abn_iu_{0}^{2}L^2}{16\hbar v_{D}^{}}\frac{\left( \varepsilon_{f}^{2}+\frac{\Delta ^2}{4} \right) i\cos\varphi +\varepsilon _{\mathbf{k}}^{+}\Delta \sin\varphi }{\varepsilon _{\mathbf{k}}^{+3}}\left[\widetilde{\mathcal{A} }_{y,\mathbf{k}}^{++}\left( \omega \right) f(\varepsilon _{\mathbf{k}}^{+})\bigg|_{\omega =0}^{}-\widetilde{\mathcal{A} }_{y,\mathbf{k}}^{++}\left( \omega \right) f(-\varepsilon _{\mathbf{k}}^{+})\bigg|_{\omega =\varepsilon _{\mathbf{k}}^{+}+\varepsilon _f>0}^{}\right],
				\\
				\chi^+_{xy,\mathbf{k},\text{v}}= &-\frac{iabn_iu_{0}^{2}L^2}{4\hbar v_{D}^{}}\frac{-\varepsilon _{\mathbf{k}}^{+}\frac{\Delta}{2}\cos\varphi -i(\frac{\Delta}{2})^2\sin\varphi}{\varepsilon _{\mathbf{k}}^{+2}}\frac{\widetilde{\mathcal{A} }_{x,\mathbf{k}}^{++}\left( \omega \right)}{\varepsilon _{\mathbf{k}}^{+}
					+\varepsilon _{f}}\bigg|_{\omega =\varepsilon _\mathbf{k}^+ - \varepsilon _f>0} \\
				&+\frac{\pi v_Dabn_iu_{0}^{2}L^2}{8\hbar ^2v_{D}^{2}}\frac{-\varepsilon _{\mathbf{k}}^{+}\frac{\Delta}{2}\cos\varphi -i(\frac{\Delta}{2})^2\sin\varphi}{\varepsilon _{\mathbf{k}}^{+3}}\widetilde{\mathcal{A} }_{y,\mathbf{k}}^{++}\left( \omega \right) f\left( \varepsilon _{\mathbf{k}}^{+} \right) \bigg|_{\omega =0}^{},\\
				\chi^+_{xy,\mathbf{k},\text{vi}}= &-\frac{i\tau abn_iu_{0}^{2}L^2}{4\hbar v_{D}^{}}\frac{\varepsilon _{\mathbf{k}}^{+}\frac{\Delta}{2}\sin\varphi +i(\frac{\Delta}{2})^2\cos\varphi}{\varepsilon _{\mathbf{k}}^{+2}}\frac{\widetilde{\mathcal{A} }_{x,\mathbf{k}}^{++}\left( \omega \right)}{\varepsilon _{\mathbf{k}}^{+}+\varepsilon _{f}}\bigg|_{\omega =\varepsilon _\mathbf{k}^+ - \varepsilon _f>0} \\
				&+\frac{\tau \pi abn_iu_{0}^{2}L^2}{8\hbar ^2v_{D}^{2}}\frac{\varepsilon _{\mathbf{k}}^{+}\frac{\Delta}{2}\sin\varphi +i(\frac{\Delta}{2})^2\cos\varphi}{\varepsilon _{\mathbf{k}}^{+2}}\widetilde{\mathcal{A} }_{y,\mathbf{k}}^{++}\left( \omega \right) f\left( \varepsilon _{\mathbf{k}}^{+} \right) \bigg|_{\omega =0}^{}.\\
			\end{aligned}
		\end{equation}
		According to the Fig. \ref{fig1}, the (vii)-(ix) diagrams are for $+$ band,
		\begin{equation}
			\begin{aligned}
				\chi^-_{xy,\mathbf{k},\text{vii}}= &\frac{ivabn_iu_{0}^{2}L^2}{8\hbar v_{D}^{}}\frac{-\varepsilon _{f}\Delta \cos\varphi +\left( \varepsilon _{f}^{2}+\frac{\Delta ^2}{4} \right) i\sin\varphi }{\varepsilon_{f}^{2}}\frac{\widetilde{\mathcal{A} }_{x,\mathbf{k}}^{++}\left( \omega \right)}{\varepsilon _{\mathbf{k}}^{+}+\varepsilon _f}\bigg|_{\omega =\varepsilon_\mathbf{k}^+-\varepsilon _f>0}
				\\
				&-\frac{\pi abn_iu_{0}^{2}L^2}{16\hbar v_{D}^{}}\frac{-\varepsilon _{\mathbf{k}}^{+}\Delta \cos\varphi +\left( \varepsilon _{\mathbf{k}}^{+2}+\frac{\Delta ^2}{4} \right) i\sin\varphi }{\varepsilon _{\mathbf{k}}^{+3}}\widetilde{\mathcal{A} }_{x,\mathbf{k}}^{++}\left( \omega \right) f(\varepsilon _\mathbf{k}^+)\bigg|_{\omega =0},\\
				\chi^-_{xy,\mathbf{k},\text{viii}}= &\frac{i\tau abn_iu_{0}^{2}L^2}{8\hbar v_{D}^{}}\frac{-\left( \varepsilon _{f}^{2}+\frac{\Delta ^2}{4} \right) i\cos\varphi +\varepsilon _{f}\Delta \sin\varphi }{\varepsilon _{f}^{2}}\frac{\widetilde{\mathcal{A} }_{y,\mathbf{k}}^{++}\left( \omega \right)}{\varepsilon _{\mathbf{k}}^{+}+\varepsilon _{f}}\bigg|_{\omega =\varepsilon_\mathbf{k}^+-\varepsilon_f >0}^{}\\
				&-\frac{\pi \tau abn_iu_{0}^{2}L^2}{16\hbar v_{D}^{}}\frac{-\left( \varepsilon _{\mathbf{k}}^{+2}+\frac{\Delta ^2}{4} \right) i\cos\varphi +\varepsilon _{\mathbf{k}}^{+}\Delta \sin\varphi }{\varepsilon _{\mathbf{k}}^{+3}}f(\varepsilon _a)\widetilde{\mathcal{A} }_{y,\mathbf{k}}^{++}\left( \omega \right) \bigg|_{\omega =0}^{},\\
				\chi^-_{xy,\mathbf{k},\text{ix}}= &\frac{iabn_iu_{0}^{2}L^2}{4\hbar v_{D}^{}}\frac{-\varepsilon _{\mathbf{k}}^{+}\frac{\Delta}{2}\cos\varphi +i(\frac{\Delta}{2})^2\sin\varphi}{\varepsilon _{\mathbf{k}}^{+2}}\frac{\widetilde{\mathcal{A} }_{x,\mathbf{k}}^{++}\left( \omega \right)}{\varepsilon _{\mathbf{k}}^{+} +\varepsilon _{f}}\bigg|_{\omega =\varepsilon _f -\varepsilon _\mathbf{k}^+ >0}\\
				&+\frac{\pi v_Dabn_iu_{0}^{2}L^2}{8\hbar ^2v_{D}^{2}}\frac{-\varepsilon _{\mathbf{k}}^{+}\frac{\Delta}{2}\cos\varphi +i(\frac{\Delta}{2})^2\sin\varphi}{\varepsilon _{\mathbf{k}}^{+3}}\widetilde{\mathcal{A} }_{x,\mathbf{k}}^{++}\left( \omega \right) f\left( \varepsilon _{\mathbf{k}}^{+} \right) \bigg|_{\omega =0}^{},\\
				\chi^-_{xy,\mathbf{k},\text{x}}= &\frac{i\tau abn_iu_{0}^{2}L^2}{4\hbar v_{D}^{}}\frac{\varepsilon _{\mathbf{k}}^{+}\frac{\Delta}{2}\sin\varphi -i(\frac{\Delta}{2})^2\cos\varphi}{\varepsilon _{\mathbf{k}}^{+2}}\frac{\widetilde{\mathcal{A} }_{x,\mathbf{k}}^{++}\left( \omega \right)}{\varepsilon _{f}+\varepsilon _{\mathbf{k}}^{+}}\bigg|_{\omega =\varepsilon _f -\varepsilon _\mathbf{k}^+ >0}\\
				&+\frac{\tau \pi abn_iu_{0}^{2}L^2}{8\hbar ^2v_{D}^{2}}\frac{\varepsilon _{\mathbf{k}}^{+}\frac{\Delta}{2}\sin\varphi -i(\frac{\Delta}{2})^2\cos\varphi}{\varepsilon _{\mathbf{k}}^{+2}}
				\widetilde{\mathcal{A} }_{y,\mathbf{k}}^{++}(\omega)f\left( \varepsilon _{\mathbf{k}}^{+} \right) \bigg|_{\omega =0}^{}.\\
			\end{aligned}
		\end{equation}
		
		$\chi_{yx,\mathbf{k}}$ is solved in the same way.
		\newpage
		
		\subsection{The dressed DC and QM}
		\label{AppendixB5}
		The diagonal dressed Berry curvature and quantum metric for $\pm$ band are written as
		\begin{equation}
			\begin{aligned}
				&g^{\text{d},+}_{xy,\mathbf{k}}/\Omega^{\text{d},+}_{xy,\mathbf{k}}=-\frac{1}{2\pi}\text{Im}\times/-\frac{1}{\pi}\text{Re}\times\\
				\frac{\sin\theta n_iu_0^2L^2}{8v_D\hbar }\Bigg\{&\frac{i}{2}\frac{- \varepsilon^{+}_{f}\frac{\Delta}{2} \cos\varphi +\varepsilon^{2}_{f}i\sin\varphi - \varepsilon_{f}\frac{\Delta}{2} \cos\varphi+(\frac{\Delta}{2})^2i\sin\varphi  }{\varepsilon^{+2}_{f}} \frac{\widetilde{\mathcal{A}}^{++}_{x,\mathbf{k}}(\omega)}{\varepsilon^{+}_\mathbf{k}+\varepsilon_f}\bigg|_{\omega=\varepsilon^{+}_\mathbf{k}-\varepsilon_{f}>0}\\
				-&\frac{ \pi}{4}\frac{- \varepsilon^+_{\mathbf{k}}\frac{\Delta}{2} \cos\varphi +\varepsilon^{+2}_{\mathbf{k}}i\sin\varphi - \varepsilon^+_{\mathbf{k}}\frac{\Delta}{2} \cos\varphi+(\frac{\Delta}{2})^2 i\sin\varphi  }{\varepsilon^{+3}_{\mathbf{k}}}f(\varepsilon^+_\mathbf{k})\widetilde{\mathcal{A}}^{++}_{x,\mathbf{k}}(\omega)\bigg|_{\omega=0}\\
				+&\frac{i\tau }{2}\frac{-\varepsilon^{2}_{f} i\cos\varphi +\varepsilon_{f}\frac{\Delta}{2} \sin\varphi+\varepsilon_{f}\frac{\Delta}{2} \sin\varphi-(\frac{\Delta}{2})^2i\cos\varphi }{\varepsilon^{+2}_{f}} \frac{ \widetilde{\mathcal{A}}^{++}_{y,\mathbf{k}}(\omega)}{\varepsilon^+_\mathbf{k}+\varepsilon_{f}}\bigg|_{\omega=\varepsilon^{+}_\mathbf{k}-\varepsilon_{f}>0} \\
				-&\frac{ \pi\tau }{4} \frac{-\varepsilon^{+2}_{\mathbf{k}}i\cos\varphi +\varepsilon^+_{\mathbf{k}}\frac{\Delta}{2} \sin\varphi+\varepsilon^+_{\mathbf{k}}\frac{\Delta}{2} \sin\varphi-(\frac{\Delta}{2})^2i\cos\varphi }{\varepsilon^{+3}_{\mathbf{k}}}f(\varepsilon^+_\mathbf{k})\widetilde{\mathcal{A}}^{++}_{y,\mathbf{k}}(\omega)\bigg|_{\omega=0}\\
				+&i\frac{- \varepsilon^+_{\mathbf{k}}\frac{\Delta}{2}\cos\varphi+i(\frac{\Delta}{2} )^2 \sin\varphi}{\varepsilon^{+2}_{\mathbf{k}}}\frac{\widetilde{\mathcal{A}}^{++}_{x,\mathbf{k}}(\omega)}{\varepsilon^{+}_{\mathbf{k}} +\varepsilon^+_f}\bigg|_{\omega=\varepsilon_{f}-\varepsilon^{+}_\mathbf{k}>0}+\frac{\pi}{2 }\frac{- \varepsilon^+_{\mathbf{k}}\frac{\Delta}{2}\cos\varphi+i(\frac{\Delta}{2})^2 \sin\varphi}{\varepsilon^{+3}_{\mathbf{k}}} f(\varepsilon^+_\mathbf{k})\widetilde{\mathcal{A}}^{++}_{x,\mathbf{k}}(\omega)\bigg|_{\omega=0}\\
				+&i\tau\frac{\varepsilon^+_{\mathbf{k}} \frac{\Delta}{2}\sin\varphi-i(\frac{\Delta}{2} )^2 \cos\varphi}{\varepsilon^{+2}_{\mathbf{k}}}\frac{\widetilde{\mathcal{A}}^{++}_{y,\mathbf{k}}(\omega)}{\varepsilon_{f} +\varepsilon^+_{\mathbf{k}}}\bigg|_{\omega=\varepsilon_{f}-\varepsilon^{+}_\mathbf{k}>0}+\frac{\tau \pi}{2 } \frac{\varepsilon^+_{\mathbf{k}} \frac{\Delta}{2}\sin\varphi-i(\frac{\Delta}{2} )^2 \cos\varphi}{\varepsilon^{+3}_{\mathbf{k}}}f(\varepsilon^+_\mathbf{k})\widetilde{\mathcal{A}}^{++}_{y,\mathbf{k}}(\omega)\bigg|_{\omega=0}\\
				\pm &\frac{i\tau}{2}\frac{ \varepsilon^{2}_{f} i\cos\varphi+\varepsilon^{+}_{f}\frac{\Delta}{2}\sin\varphi  + (\frac{\Delta}{2} )^2 i\cos\varphi +\varepsilon_{f}\frac{\Delta}{2}\sin\varphi }{\varepsilon^{+2}_{f}} \frac{\widetilde{\mathcal{A}}^{++}_{y,\mathbf{k}}(\omega)}{\varepsilon^{+}_\mathbf{k}+\varepsilon_f}\bigg|_{\omega=\varepsilon^{+}_\mathbf{k}-\varepsilon_{f}>0}\\
				\mp &\frac{ \pi\tau}{4}\frac{\varepsilon^{+2}_{\mathbf{k}} i\cos\varphi+\varepsilon^{+}_{\mathbf{k}}\frac{\Delta}{2}\sin\varphi  + (\frac{\Delta}{2} )^2 i\cos\varphi +\varepsilon^{+}_{\mathbf{k}}\frac{\Delta}{2}\sin\varphi  }{\varepsilon^{+3}_{\mathbf{k}}}f(\varepsilon^+_\mathbf{k})\widetilde{\mathcal{A}}^{++}_{y,\mathbf{k}}(\omega)\bigg|_{\omega=0}\\
				\pm &\frac{i }{2}\frac{-\varepsilon^{2}_{f} i\sin\varphi-\varepsilon_{f}\frac{\Delta}{2}\cos\varphi -\varepsilon_{f}\frac{\Delta}{2}\cos\varphi -(\frac{\Delta}{2})^2 i\sin\varphi }{\varepsilon^{+2}_{f}} \frac{\widetilde{\mathcal{A}}^{++}_{x,\mathbf{k}}(\omega)}{\varepsilon^+_\mathbf{k}+\varepsilon_{f}}\bigg|_{\omega=\varepsilon^{+}_\mathbf{k}-\varepsilon_{f}>0} \\
				\mp &\frac{ \pi }{4} \frac{-\varepsilon^{+2}_{\mathbf{k}} i\sin\varphi-\varepsilon^+_{\mathbf{k}}\frac{\Delta}{2}\cos\varphi -\varepsilon^{+}_{\mathbf{k}}\frac{\Delta}{2}\cos\varphi -(\frac{\Delta}{2})^2 i\sin\varphi }{\varepsilon^{+3}_{\mathbf{k}}}f(\varepsilon^+_\mathbf{k})\widetilde{\mathcal{A}}^{++}_{x,\mathbf{k}}(\omega)\bigg|_{\omega=0}\\
				\pm &i \tau\frac{ \varepsilon^+_{\mathbf{k}}\frac{\Delta}{2}\sin\varphi+i(\frac{\Delta}{2} )^2 \cos\varphi}{\varepsilon^{+2}_{\mathbf{k}}}\frac{ \widetilde{\mathcal{A}}^{++}_{y,\mathbf{k}}(\omega)}{\varepsilon^{+}_{\mathbf{k}} +\varepsilon^+_f}\bigg|_{\omega=\varepsilon_{f}-\varepsilon^{+}_\mathbf{k}>0} +\frac{\tau \pi }{2}\frac{ \varepsilon^+_{\mathbf{k}}\frac{\Delta}{2}\sin\varphi+i(\frac{\Delta}{2})^2 \cos\varphi}{\varepsilon^{+3}_{\mathbf{k}}} f(\varepsilon^+_\mathbf{k})\widetilde{\mathcal{A}}^{++}_{y,\mathbf{k}}(\omega)\bigg|_{\omega=0}\\
				\pm &i\frac{-\varepsilon^+_{\mathbf{k}} \frac{\Delta}{2}\cos\varphi-i(\frac{\Delta}{2} )^2 \sin\varphi}{\varepsilon^{+2}_{\mathbf{k}}}\frac{\widetilde{\mathcal{A}}^{++}_{x,\mathbf{k}}(\omega)}{\varepsilon_{f} +\varepsilon^+_{\mathbf{k}}}\bigg|_{\omega=\varepsilon_{f}-\varepsilon^{+}_\mathbf{k}>0}+\frac{ \pi }{2} \frac{-\varepsilon^+_{\mathbf{k}} \frac{\Delta}{2}\cos\varphi-i(\frac{\Delta}{2} )^2 \sin\varphi}{\varepsilon^{+3}_{\mathbf{k}}}f(\varepsilon^+_\mathbf{k})\widetilde{\mathcal{A}}^{++}_{x,\mathbf{k}}(\omega)\bigg|_{\omega=0}\Bigg\}.\\
			\end{aligned}
		\end{equation}
		\newpage
		\begin{equation}
			\begin{aligned}
				&g^{\text{d},-}_{xy,\mathbf{k}}/\Omega^{\text{d},-}_{xy,\mathbf{k}}=-\frac{1}{2\pi}\text{Im}\times/-\frac{1}{\pi}\text{Re}\times\\
				\frac{\sin\theta n_iu_0^2L^2}{8v_D\hbar}\Bigg\{&-\frac{i }{2}\frac{- \varepsilon_{f} \frac{\Delta}{2} \cos\varphi -\varepsilon^{+2}_{f}i\sin\varphi - \varepsilon_{f}\frac{\Delta}{2} \cos\varphi-(\frac{\Delta}{2})^2 i\sin\varphi  }{\varepsilon^{+2}_{f}}\frac{\widetilde{\mathcal{A}}^{++}_{x,\mathbf{k}}(\omega)}{\varepsilon^+_\mathbf{k}+\varepsilon_f}\bigg|_{\omega=\varepsilon_{f}-\varepsilon^{+}_\mathbf{k}>0}\\
				&-\frac{ \pi }{4}\frac{- \varepsilon^+_{\mathbf{k}}\frac{\Delta}{2} \cos\varphi -\varepsilon^{+2}_{\mathbf{k}}i\sin\varphi - \varepsilon^+_{\mathbf{k}}\frac{\Delta}{2} \cos\varphi-(\frac{\Delta}{2})^2 i\sin\varphi  }{\varepsilon^{+3}_{\mathbf{k}}}\widetilde{\mathcal{A}}^{++}_{x,\mathbf{k}}(\omega)\left[f(\varepsilon^{+}_\mathbf{k})\bigg|_{\omega=0}-f(-\varepsilon^{+}_\mathbf{k})\bigg|_{\omega=\varepsilon^{+}_\mathbf{k}+\varepsilon_f}\right]\\
				&-\frac{ i\tau }{2}\frac{\varepsilon^{+2}_{f}i\cos\varphi +\varepsilon_{f}\frac{\Delta}{2} \sin\varphi+\varepsilon_{f}\frac{\Delta}{2} \sin\varphi+(\frac{\Delta}{2})^2i\cos\varphi }{\varepsilon^{+2}_{f}} \frac{\widetilde{\mathcal{A}}^{++}_{y,\mathbf{k}}(\omega)}{\varepsilon^+_\mathbf{k}+\varepsilon_f}\bigg|_{\omega=\varepsilon_{f}-\varepsilon^{+}_\mathbf{k}>0}\\
				&-\frac{ \pi\tau }{4}\frac{\varepsilon^{+2}_{\mathbf{k}}i\cos\varphi +\varepsilon^+_{\mathbf{k}}\frac{\Delta}{2} \sin\varphi+\varepsilon^+_{\mathbf{k}}\frac{\Delta}{2} \sin\varphi+(\frac{\Delta}{2})^2i\cos\varphi }{\varepsilon^{+3}_{\mathbf{k}}}\widetilde{\mathcal{A}}^{++}_{y,\mathbf{k}}(\omega)\left[f(\varepsilon^{+}_\mathbf{k})\bigg|_{\omega=0}-f(-\varepsilon^{+}_\mathbf{k})\bigg|_{\omega=\varepsilon^{+}_\mathbf{k}+\varepsilon_f}\right]\\
				&-i \frac{- \varepsilon^+_{\mathbf{k}}\frac{\Delta}{2}\cos\varphi-i(\frac{\Delta}{2})^2 \sin\varphi}{\varepsilon^{+2}_{\mathbf{k}}}\frac{\widetilde{\mathcal{A}}^{++}_{x,\mathbf{k}}(\omega)}{\varepsilon^+_{\mathbf{k}}+\varepsilon_{f}}\bigg|_{\omega=\varepsilon^{+}_\mathbf{k}-\varepsilon_{f}>0}+\frac{ \pi }{2} \frac{- \varepsilon^+_{\mathbf{k}}\frac{\Delta}{2}\cos\varphi-i(\frac{\Delta}{2} )^2 \sin\varphi}{\varepsilon^{+3}_{\mathbf{k}}}f(\varepsilon^+_\mathbf{k})\widetilde{\mathcal{A}}^{++}_{x,\mathbf{k}}(\omega)\bigg|_{\omega=0}\\
				&-i\tau\frac{\varepsilon^+_{\mathbf{k}}\frac{\Delta}{2}\sin\varphi+i(\frac{\Delta}{2} )^2 \cos\varphi}{\varepsilon^{+2}_{\mathbf{k}}}\frac{\widetilde{\mathcal{A}}^{++}_{y,\mathbf{k}}(\omega)}{\varepsilon_{f}+\varepsilon^+_{\mathbf{k}}}\bigg|_{\omega=\varepsilon^{+}_\mathbf{k}-\varepsilon_{f}>0}+\frac{\tau \pi }{2}\frac{\varepsilon^+_{\mathbf{k}}\frac{\Delta}{2}\sin\varphi+i(\frac{\Delta}{2} )^2 \cos\varphi}{\varepsilon^{+3}_{\mathbf{k}}}f(\varepsilon^+_\mathbf{k})\widetilde{\mathcal{A}}^{++}_{y,\mathbf{k}}(\omega)\bigg|_{\omega=0}\\
				&\mp \frac{i\tau }{2}\frac{- \varepsilon^{+2}_{f} i\cos\varphi+\varepsilon_{f}\frac{\Delta}{2} \sin\varphi - (\frac{\Delta}{2})^2 i\cos\varphi +\varepsilon_{f}\frac{\Delta}{2}\sin\varphi }{\varepsilon^{+2}_{f}}\frac{\widetilde{\mathcal{A}}^{++}_{y,\mathbf{k}}(\omega)}{\varepsilon^+_\mathbf{k}+\varepsilon_f}\bigg|_{\omega=\varepsilon_{f}-\varepsilon^{+}_\mathbf{k}>0}\\
				&\mp \frac{ \pi \tau}{4}\frac{- \varepsilon^{+2}_{\mathbf{k}} i\cos\varphi+\varepsilon^{+}_{\mathbf{k}}\frac{\Delta}{2} \sin\varphi - (\frac{\Delta}{2})^2 i\cos\varphi +\varepsilon^{+}_{\mathbf{k}}\frac{\Delta}{2}\sin\varphi }{\varepsilon^{+3}_{\mathbf{k}}}\widetilde{\mathcal{A}}^{++}_{y,\mathbf{k}}(\omega)\left[f(\varepsilon^{+}_\mathbf{k})\bigg|_{\omega=0}-f(-\varepsilon^{+}_\mathbf{k})\bigg|_{\omega=\varepsilon^{+}_\mathbf{k}+\varepsilon_f}\right]\\
				&\mp \frac{ i }{2}\frac{\varepsilon^{+2}_{f} i\sin\varphi-\varepsilon_{f}\frac{\Delta}{2}\cos\varphi -\varepsilon_{f}\frac{\Delta}{2}\cos\varphi +(\frac{\Delta}{2})^2 i\sin\varphi }{\varepsilon^{+2}_{f}} \frac{\widetilde{\mathcal{A}}^{++}_{x,\mathbf{k}}(\omega)}{\varepsilon^+_\mathbf{k}+\varepsilon_f}\bigg|_{\omega=\varepsilon_{f}-\varepsilon^{+}_\mathbf{k}>0}\\
				&\mp \frac{ \pi }{4}\frac{\varepsilon^{+2}_{\mathbf{k}} i\sin\varphi-\varepsilon^{+}_{\mathbf{k}}\frac{\Delta}{2}\cos\varphi -\varepsilon^{+}_{\mathbf{k}}\frac{\Delta}{2}\cos\varphi +(\frac{\Delta}{2})^2 i\sin\varphi }{\varepsilon^{+3}_{\mathbf{k}}}\widetilde{\mathcal{A}}^{++}_{x,\mathbf{k}}(\omega)\left[f(\varepsilon^{+}_\mathbf{k})\bigg|_{\omega=0}-f(-\varepsilon^{+}_\mathbf{k})\bigg|_{\omega=\varepsilon^{+}_\mathbf{k}+\varepsilon_f}\right]\\
				&\mp i\tau  \frac{ \varepsilon^+_{\mathbf{k}}\frac{\Delta}{2}\sin\varphi-i(\frac{\Delta}{2})^2 \cos\varphi}{\varepsilon^{+2}_{\mathbf{k}}}\frac{\widetilde{\mathcal{A}}^{++}_{y,\mathbf{k}}(\omega)}{\varepsilon^+_{\mathbf{k}}+\varepsilon_{f}}\bigg|_{\omega=\varepsilon^{+}_\mathbf{k}-\varepsilon_{f}>0} +\frac{ \pi \tau }{2}\frac{ \varepsilon^+_{\mathbf{k}}\frac{\Delta}{2}\sin\varphi-i(\frac{\Delta}{2} )^2 \cos\varphi}{\varepsilon^{+3}_{\mathbf{k}}}f(\varepsilon^+_\mathbf{k})\widetilde{\mathcal{A}}^{++}_{y,\mathbf{k}}(\omega)\bigg|_{\omega=0}\\
				&\mp i\frac{-\varepsilon^+_{\mathbf{k}}\frac{\Delta}{2}\cos\varphi+i(\frac{\Delta}{2} )^2 \sin\varphi}{\varepsilon^{+2}_{\mathbf{k}}}\frac{\widetilde{\mathcal{A}}^{++}_{x,\mathbf{k}}(\omega)}{\varepsilon_{f}+\varepsilon^+_{\mathbf{k}}}\bigg|_{\omega=\varepsilon^{+}_\mathbf{k}-\varepsilon_{f}>0} +\frac{ \pi }{2}\frac{-\varepsilon^+_{\mathbf{k}}\frac{\Delta}{2}\cos\varphi+i(\frac{\Delta}{2} )^2 \sin\varphi}{\varepsilon^{+3}_{\mathbf{k}}}f(\varepsilon^+_\mathbf{k})\widetilde{\mathcal{A}}^{++}_{x,\mathbf{k}}(\omega)\bigg|_{\omega=0}\Bigg\}.\\
			\end{aligned}
		\end{equation}
		
	\end{widetext}

	%\bibliography{ref}
	
	%\bibliography{1}
	

\end{document}